\documentclass[lettersize,journal]{IEEEtran}
\usepackage{algorithmicx,algorithm}
\usepackage{amsmath}
\usepackage{amsthm}
\usepackage{amsfonts}
\usepackage{subfigure}
\usepackage{mathrsfs}
\usepackage{pifont}
\usepackage{amssymb}
\usepackage{verbatim}
\usepackage{upgreek}
\usepackage{color}
\usepackage{epsfig}
\usepackage{booktabs}
\usepackage{bm}
\usepackage{setspace}

\usepackage[hyphens]{url}

\usepackage[latin1]{inputenc}
\usepackage{graphicx}
\usepackage{caption}

\usepackage[overload]{empheq}
\usepackage{fancyhdr}
\usepackage[colorlinks]{hyperref}

\usepackage{cite}

\usepackage{threeparttable}

\usepackage{multirow}
\usepackage{array}

\makeatletter
\renewcommand{\maketag@@@}[1]{\hbox{\m@th\normalsize\normalfont#1}}%
\makeatother

\newtheorem{theorem}{Theorem}[section]

\newtheorem{lemma}[theorem]{Lemma}

\title{Hybrid RIS-Enhanced ISAC Secure Systems: Joint Optimization in the Presence of an Extended Target}

\author{Yu Yao, \IEEEmembership{Member, IEEE,}
              Junhao Zhang, 
              Pu Miao, \IEEEmembership{Member, IEEE,}
              Long Zhang, \IEEEmembership{Member, IEEE,}
              Gaojie Chen, \IEEEmembership{Senior Member, IEEE,}
              Feng Shu, \IEEEmembership{Member, IEEE,}
            and Kai-Kit Wong, \IEEEmembership{Fellow, IEEE}
\thanks{Yu Yao, Junhao Zhang and Feng Shu are with the School of Information and Communication Engineering, Hainan University, Haikou, 570228, China (e-mails: yaoyu@hainanu.edu.cn; capablezzz316@163.com; shufeng0101@163.com).}
\thanks{Pu Miao is with the School of Electronic and Information Engineering, Qingdao University, Qingdao, 266000, China (e-mail: mpvae@qdu.edu.cn).}
\thanks{Long Zhang and Gaojie Chen are with the School of Flexible Electronics (SoFE), Sun Yat-sen University, Shenzhen, Guangdong 518107, China (e-mails: zhangl02@pcl.ac.cn; gaojie.chen@ieee.org).}
\thanks{Kai-Kit Wong is with the Department of Electronic and Electrical Engineering, University College London, London WC1E 6BT, U.K (e-mail: kai-kit.wong@ucl.ac.uk). He is also affiliated with Yonsei Frontier Lab, Yonsei University, Seoul 03722, South Korea.}
}
\begin{document}

\maketitle

\begin{abstract}

Unlike the conventional fully-passive and fully-active reconfigurable intelligent surfaces (RISs), a hybrid RIS consisting of active and passive reflection units has recently been concerned, which can exploit their integrated advantages to alleviate the RIS-induced path loss.
In this paper, we investigate a novel security strategy where the multiple hybrid RIS-aided integrated sensing and communication (ISAC) system communicates with downlink users and senses an extended target synchronously.
{\color{blue}Assuming imperfectly known target location (TL), we consider the joint design of the transmit signal and receive filter bank of the base station (BS), the receive beamformers of all users
 and the weights of the hybrid RIS.}
An optimization problem is formulated for maximizing {\color{blue}the worst-case sensing} signal-to-interference-plus-noise-ratio (SINR) subject to secure communication and system power budget constraints.
To address this non-convex problem, we leverage {\color{blue}generalized fractional programming (GFP)} and penalty-dual-decomposition (PDD), and propose a security solution that efficiently optimizes all variables by employing convex optimization approaches.
Simulation results show that by incorporating the multiple hybrid RIS into the optimization design, the extended target detection and secure transmission performance of ISAC systems are improved over the state-of-the-art RIS-aided ISAC approaches.

\end{abstract}


\section{Introduction}

The forthcoming sixth-generation (6G) wireless systems will enable many promising applications for instance vehicle-to-everything (V2X), smart cities and extended reality (XR)~\cite{2020Andrew}.
Therefore, there is a critical need to integrate environment sensing functionality into communication systems to yield perceptive wireless networks through the concept of integrated sensing and communications (ISAC)~\cite{2021Fan}.
{\color{blue}However, ISAC systems are confronted with serious security problems owing to not only the broadcasting nature of wireless transmissions but also the fact that the detected target might possibly eavesdrop
the confidential information-bearing signal sent to the licensed receivers within range, which poses a new security concern~\cite{Chen3, Chen2, 2023Jia,2023Ren}.}
%In ISAC systems, the dual-functional transmit signal is custom designed so as to achieve dual goals of radar sensing and information transmission, which brings potential security threats~\cite{ChenGLOBECOM}.
%As the transmit waveform carries private information for the communication receivers, the radar target, as a possible eavesdropper (Eve), could easily intercept the communication information intended for authorized receivers
%For this purpose, physical layer security (PLS) schemes could be used to perform the dual task in security-critical ISAC designs.

To overcome the potential security problem in ISAC networks, in~\cite{2018Jonathon}, sensing-centric ISAC systems were devised to convey a mixture of two different waveforms, containing communication information for the authorized receivers and artificial noise (AN) to disturb the Eve, both of which can be employed for radar target detection.
{\color{blue}However, the extra AN would unavoidably induce interference and deteriorate the network performance.}
Furthermore, communication-centric ISAC systems face severe security threats, since transmit waveforms, enabled for target detection, are usually vulnerable to wiretapping attacks owing to the reuse of a data embedding waveform~\cite{2022Su,2022Xu}.
An optimization problem was formulated to maximize the radar received signal-to-interference-plus-noise ratio (SINR) with a particular secrecy rate requirement~\cite{2017Batu}.
Different from the secrecy rate based approaches, the directional modulation (DM) scheme adjusts the amplitude and phase of the symbols at the desired receivers while disturbing the symbols in unintended directions, which means that the modulation occurs at the antenna level instead of at the baseband level.
The DM-based technique is based on the principles of utilizing constructive interference (CI)~\cite{2018LiuTSP, 2019Khandaker}, where the received waveform is not required to be aligned with the desired symbols, but is kept away from the detection threshold levels of the constellation.
{\color{blue}The symbol-level precoding scheme, which exploits more available degrees of freedom (DoFs) in both the spatial and temporal domains, can offer enhanced transmit beampatterns to realize superior sensing performance, as well as convert harmful multiuser interference (MUI) into CI for enhancing multi-user communication quality~\cite{2022Su}.
However, adding sensing functionality to communication network degrades the communication quality since the two objectives might conflict with each other.}

%It is demonstrable that the CI precoding scheme is beneficial to improve the secure communication performance.
%The symbol-level precoding method was used to optimize the dual-functional waveform for utilizing the intrinsic multiuser interference (MUI) as beneficial power to secure ISAC systems~\cite{2022Su}.
%Furthermore, the authors of [5], [6] employed the symbol-level precoding method to design the transmit waveform for exploiting the inherent multiuser interference to further secure ISAC systems.
%The authors of [5], [6] employed the symbol-level precoding technique to design the signal waveform for exploiting the inherent multiuser interference to further secure ISAC systems.
%Assuming that the radar target serves as an attacker damaging the wireless network security, the authors of~\cite{2017Kalantari} exploited the AN scheme to secure an ISAC system against wiretapping.
%The DM method was used to optimize the radar received SINR under transmit power and information secrecy requirements.
%Especially, the AN and CI  methods are jointly used to optimize security performance under the hypotheses of imperfect channel state information (CSI)~\cite{2019Khandaker}, and shown to outperform the traditional AN-assisted security design.
%On the other hand, it is common knowledge that

Reconfigurable intelligent surfaces (RISs) are capable of adjusting channels to perform secure and reliable communications through designed waveform reflection which can be viewed as a promising technique for future 6G networks~\cite{2021Wu, 2021Wang}.
Inspired by the huge potential provided by a RIS, various works have been devoted to the optimization of RIS-enabled ISAC systems~\cite{2023Hua, 2022LiuRang,2022Wang, 2022Huang}.
%A RIS-enabled secure communications and sensing system was developed in~\cite{2022Du}.
Since passive RIS units only adjust the signal phase, the fully passive RIS enhanced channel suffers serious propagation attenuation, particularly when the RIS doesn't consist of a large number of reflection units.
The fully active RIS is capable of utilizing amplifiers to tackle the serious channel attenuation.
However, it requires extra power consumption to improve the strength of the reflected waveforms and noise at the active RIS~\cite{Jiangzhou2024}.
To overcome the issue, a hybrid RIS including a certain number of active units was proposed~\cite{HybridAP1,HybridAP2}, which serves two purposes of amplifying and reflecting incident signals and achieves a significant ISAC performance improvement.
A hybrid RIS can be considered as a promising framework to reduce the influences on system performance improvement that occur owing to severe transmission attenuation.
The robust beamforming optimization for a hybrid RIS-enabled ISAC system was proposed in~\cite{HybridAP1}.
The work of~\cite{HybridAP2} developed a design scheme to optimize the data rate for a hybrid RIS-assisted communication network that can operate in time-varying Rician fading channels.
The influence of the mode switching of active-passive reflection elements on the mobile edge computing-aided system was investigated in~\cite{MEC2024}.
%The joint optimization of the hybrid precoders and the hybrid RIS weights for optimizing the transmission data rate under the overall system power constraint, was proposed in~\cite{MEC2024}.
%The authors of~\cite{Chen2021} also considered adding active reflection units to passive RIS and then employed the hybrid architecture for wireless communication, and verified that the flexible RIS architecture can provide superior performance than a fully-passive RIS.
{\color{blue}A beamforming optimization method was proposed in the multi-RIS aided communication system, where the weights of multi-RIS and transmit beamforming are cooperatively designed to enhance the system performance~\cite{MultiRISImperfect023}.
Also, a similar problem was considered for multiple hybrid RIS-aided systems, which design the transmit beamforming at the base station (BS) and the reflect beamforming at the RIS to improve robustness against blockages~\cite{MultiRIS2}.}

The incorporation of a RIS into the DM technique has been revealed to be valuable for security communications~\cite{2022Shu,2021Shu,2023Wang}.
To minimize the energy consumption in DM, a RIS-assisted secure DM approach is proposed in~\cite{2022Shu}, where a design problem is established to maximize the secrecy throughput by designing the hybrid precoding and the RIS beamforming.
%With the help of RIS, DM-based communication systems are capable of transmitting two bit streams simultaneously, and based on~\cite{2021Shu}, the DM-based technique can significantly enhance the system secrecy performance.
%To deceive the potential Eve, the work of~\cite{2022Su} was presented to control the private information eavesdropped at the malicious target by enforcing the signal to fall into the destructive area.
The authors of~\cite{2023Wang} considered a joint design problem that optimizes the transmit waveform of ISAC and the weights of the RIS.
With the DM-based scheme, the transmit waveform is optimized to fall into the constructive area of potential Eves.
{\color{blue}However, the accurate prior information of extended radar target (the target aspect angle and location) is generally unknown since the angle estimation of extended radar target includes a degree of
uncertainty~\cite{MultiRISImperfect023, ExtendedTargets2024}.
The authors of~\cite{ISACCSI1} considered robust transceiver design for a secure ISAC system, and proposed robust optimization scheme with imperfect sensing channel state information (CSI).
To the best of our knowledge, the DM-based secure scheme for multi-antenna communication users in a hybrid RIS-aided ISAC system under imperfect CSI has not been studied before.}

Motivated by the aforementioned observations, this work proposes a secure optimization framework for hybrid RIS-aided ISAC systems, {\color{blue}with a particular emphasis on the maximization of worst-case SINR for extended target detection under imperfect CSI.
Specifically, considering that the radar target might intercept the confidential information intended to the secure communication users (SCUs), the DM-based method is first adopted for securing multiple hybrid RIS-aided ISAC.
The comparison between this scheme and other relevant works is provided in Table~\ref{table1} and the primary contributions of our work are listed as follows:}
%Supposing the target impulse response (TIR) and Eve's CSI are imperfectly known according to the previous observations, we propose a receive filter array with each branch regulated to a given TIR, and consider the average SINR as figure of merit.
%Different from a secrecy rate threshold constraint, we impose a CI constraint on the SCUs and a destructive interference (DI) security constraint on the Eve.
%The main contributions of this work can be summarized as follows.

\begin{table*}[t]
\centering
\footnotesize
\caption{CONTRIBUTIONS IN CONTRAST TO THE STATE-OF-THE-ART} % 表格标题
\label{table1}
\begin{tabular}{|c|c|c|c|c|c|c|}
\hline
\textbf{Related works} & ISAC & DM & Extended Target  & Imperfect CSI & Hybrid RIS & Multiple RIS \\
\hline
\cite{HybridAP1}        &\checkmark  &  &    &   & \checkmark  &    \\
\hline
\cite{MultiRISImperfect023} &  &  &    & \checkmark &  & \checkmark \\
\hline
\cite{2022Shu}             &  &    \checkmark  &    &    &     &    \\
\hline
\cite{2023Wang}        &\checkmark  &\checkmark   &    &  &     & \\
\hline
\cite{ExtendedTargets2024} &  \checkmark  &  &  \checkmark &    &    &    \\
\hline
\cite{ISACCSI1}  & \checkmark  &   &    &  \checkmark  &   &   \\
\hline
This work & \checkmark & \checkmark & \checkmark   & \checkmark &  \checkmark  & \checkmark \\
\hline
\end{tabular}
\end{table*}

\begin{itemize}
\item {\color{blue}We propose a novel multiple hybrid RIS-aided ISAC system which serves multi-antenna SCUs and detects a single-antenna Eve simultaneously.
The system model is extended to a general scenario where the target location (TL) is uncertain to the BS.}
As a consideration on secure transmission, the MUI is devised to be constructive to SCUs, while destructive to the Eve.

\item {\color{blue}To maximize the worst-case sensing SINR among all possible TLs, we jointly design the transmit signal and receive filters at the BS, the receive beamformers at each SCU, together with the weights at the hybrid RIS, while guaranteeing the CI-type communication QoS and DI-type security requirements, the RIS weight constraint, and total power budget constraint.}

 \item We utilize a sequential optimization framework based on {\color{blue}generalized fractional programming (GFP)} to handle the coupling problem of all variables.
Particularly, to solve the challenging hybrid RIS configuration problem, we appropriately split variables and leverage the penalty dual decomposition (PDD) technique combined with the penalty convex-concave procedure (PCCP) method.
We find an optimized solution to resolve the quartic optimization by tackling several quadratic sub-problems.

\item Numerical results indicate that the developed design scheme can realize better performance than the state-of-the-art benchmark methods.
Moreover, the potential of a multiple hybrid RIS for enhancing the ISAC performance is confirmed and the capability enhancement enlarges with the increasing total power consumption of the system and number of hybrid RIS reflection units.

\end{itemize}

The remainder of this paper is organized as follows:
Section~\ref{system model} introduces the hybrid RIS-enhanced ISAC secure system model and the formulation of the joint design problem.
Section~\ref{ProbForm} develops the efficient solution technique for the employed performance metric.
In Section~\ref{simulation}, simulation results are given, whereas in Section~\ref{Conclusion}, the conclusions are provided.

\emph{Notations}: Matrices are denoted by bold uppercase letters and vectors are denoted by bold lowercase letters.
The trace and the vectorization operations are represented by $\text{tr}\left( \cdot  \right)$ and $\text{vec}\left( \cdot  \right)$, respectively.
${{\left( . \right)}^{T}}$, ${{\left( . \right)}^{H}}$ and ${{\left( . \right)}^{*}}$ signify transpose, Hermitian transpose and complex conjugate of matrices, respectively.
$\otimes$ and $\circ$ are the Kronecker product and Hadamard product, respectively.
The magnitude of scalar $a$, the norm of vector $\mathbf{a}$ and the Frobenius norm of matrix $\mathbf{A}$ are signified by $\left| a \right|$ , $\left\| \mathbf{a} \right\|$, and ${{\left\| \mathbf{A} \right\|}_{F}}$, respectively.
$\mathbf{e}_{n}$ is the $n$th column of the identity matrix ${{\mathbf{I}}_{N}}$.
$\mathbb{E}\left( \cdot  \right)$ indicates the statistical expectation operation.

\section{System model}\label{system model}

\begin{figure}[ht]
\captionsetup{font={footnotesize}}
	\begin{center}
	{\includegraphics[width=7.5cm]{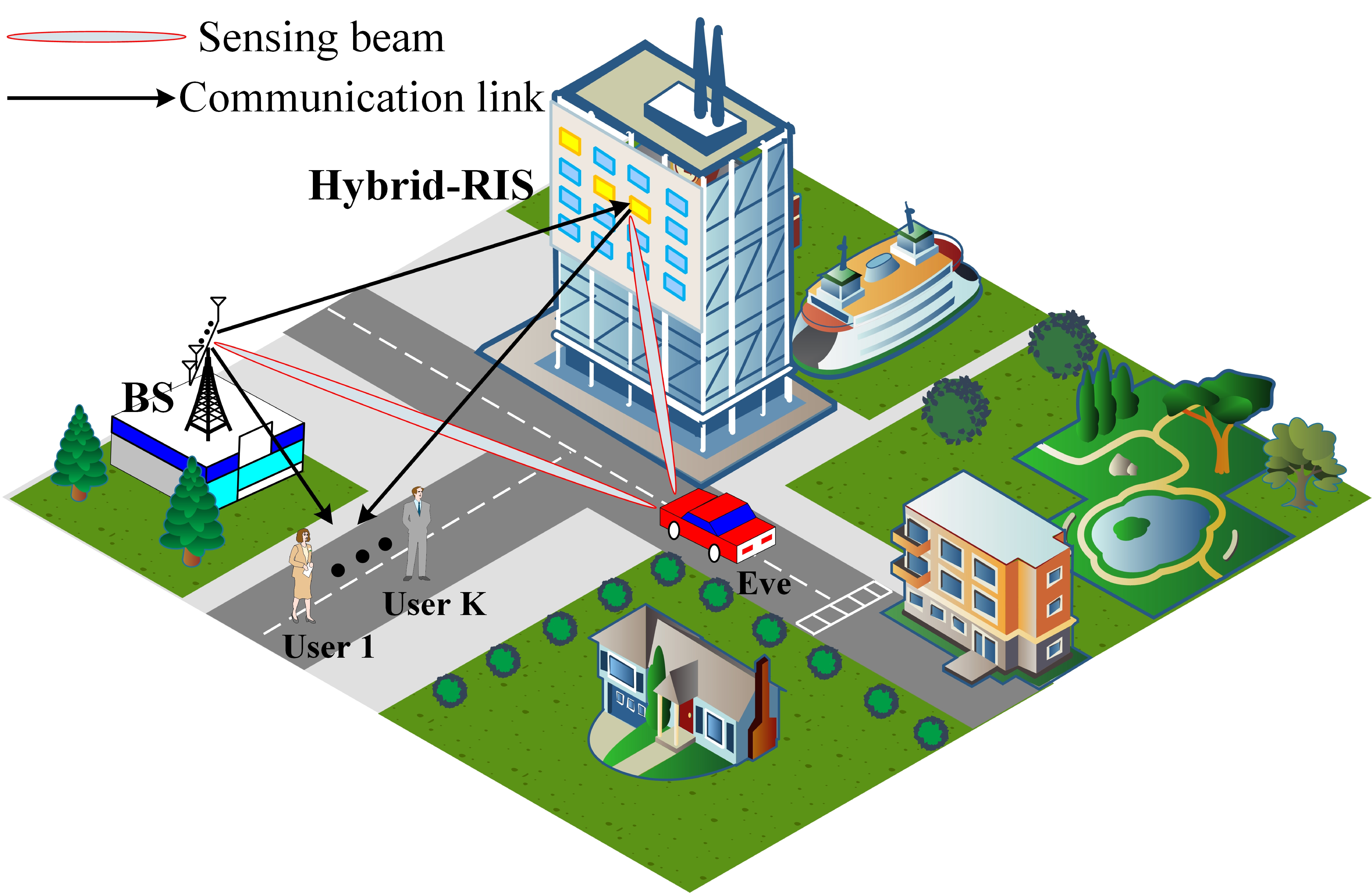}}
		\caption{A description of the studied hybrid RIS-enhanced ISAC secure system.}
		\label{HRRIS2}
	\end{center}
\end{figure}

We consider {\color{blue}multiple} hybrid RIS-aided ISAC system as described in Fig.~\ref{HRRIS2}, where a BS communicates with $K$ {\color{blue}multi-antenna} SCUs and detects an extended target with the help of {\color{blue}multiple} hybrid RIS concurrently.
The BS is equipped with ${{N}_{a}}\left( {{N}_{a}}\ge K \right)$ transmit/receive antennas.
The extended target is viewed as a single-antenna Eve which may wiretap the private message transmitted from the {\color{blue}BS} to each SCU.
Specifically, the {\color{blue}BS} firstly transmits the confidential signals to each SCU while radiating the Eve.
Then, the {\color{blue}BS} observes the scattering behavior of the extended target using the return signals reflected from the target.
Since employing traditional communication waveforms to sense a radar target needs minor modifications to developed transmission networks and does not generate additional self-interference, according to~\cite{2022LiuRang, 2022Chen}, we adopt communication waveforms to perform ISAC operation.
{\color{blue}The hybrid RIS is equipped with $N_I$ units, containing $A$ active and $N_I-A$ passive reflecting units.
We now introduce notations to mathematically model the multiple hybrid RIS.
Specifically, let $\mathbb{Q}$ be the index set of the $A$ active units, $\mathbb{Q}\in \{1,...,A\}$, and let ${{\vartheta }_{s,n}}$ be the reflection coefficient of the $n$th unit of the $s$th RIS, which is expressed as
\begin{equation}\label{HRC}
{{\vartheta }_{s,n}}=\left\{ \begin{matrix}
   \left| {{a}_{s,n}} \right|{{e}^{j{{\mu }_{s,n}}}},n\in \mathbb{Q},  \\
   {{e}^{j{{\mu }_{s,n}}}}, n\notin \mathbb{Q},  \\
\end{matrix} \right.
\end{equation}
where ${{a}_{s,n}}$ denotes the amplitude of the signal generated by the active unit, and ${{\mu }_{s,n}}\in \left[ 0,2\pi  \right)$ is the phase of the signal.
Let $\mathbf{\Theta }_s=\text{diag}\left\{ {{\vartheta }_{s,1}}, \cdots, {\vartheta }_{s,{N}_{I}} \right\}\in {{\mathbb{C}}^{{{N}_{I}}\times {{N}_{I}}}}$ be the reflection coefficient matrix of the $s$th hybrid RIS.
We define an additive decomposition for it as $\mathbf{\Theta }_s=\mathbf{\Psi }_s+\mathbf{\Phi }_s$, where $\mathbf{\Psi }_s=\mathcal{I}_{{{N}_{I}}}^{\mathbb{Q}}\circ \mathbf{\Theta }_s=\text{diag}\left\{ {{\varphi }_{s,1}}, \cdots, {\varphi }_{s,{N}_{I}} \right\}$ and $\mathbf{\Phi }_s=\left( \mathbf{I}_{{{N}_{I}}}-\mathcal{I}_{{{N}_{I}}}^{\mathbb{Q}} \right)\circ \mathbf{\Theta }_s=\text{diag}\left\{ {{\phi }_{s,1}}, \cdots, {\phi }_{s,{N}_{I}} \right\}$ contain the active and passive RIS coefficients, respectively.
where $\mathcal{I}_{{{N}_{I}}}^{\mathbb{Q}}$ is an $N_I \times N_I$ diagonal matrix whose non-zero elements are all unity and have positions determined by $\mathbb{Q}$.
The discrete RIS possesses $b$-bit controllability, where each individual reflector of the RIS is encoded by the controller to enable $2^b$ distinct phase shift modes ${{\mu }_{s,n}}=\frac{2\pi n}{{{2}^{b}}},n\in \left\{ 0,1,...,{{2}^{b}}-1 \right\},n\in {{N}_{I}}$.
We assume BS and RIS adopt uniform planner arrays (UPA) to support secure ISAC scheme.

\subsection{Multiuser Communications Model}

For UPA with ${N}_{a}=L_{x,\text{B}}\times L_{z,\text{B}}$ and ${N}_{I}=L_{x,s}\times L_{z,s}$ antennas,
the channel gain from the BS to the $s$th RIS $\mathbf{G}_s\in \mathbb{C}^{{N}_{I}\times{N}_{a}}$ is modeled as
\begin{equation}\label{RSUIRS}
{\mathbf{G}_s}=\sqrt{\frac{{\lambda }_{\text{B},s}}{1+{{\lambda }_{\text{B},s}}}}\mathbf{H}_{\text{B},s}^{\text{LoS}}
+\sqrt{\frac{1}{1+{{\lambda }_{\text{B},s}}}}\mathbf{H}_{\text{B},s}^{\text{NLoS}},
\end{equation}
where ${{\lambda }_{\text{B},s}}$ is the Rician factor of the $s$th BS-RIS link.
The line-of-sight (LoS) component $\mathbf{H}_{\text{B},s}^{\text{LoS}}\in {{\mathbb{C}}^{ N_I \times {N}_{a}}}$ and scattering component
$\mathbf{H}_{\text{B},s}^{\text{NLoS}}\in {{\mathbb{C}}^{  {N}_{I} \times {N}_{a} }}$ are respectively expressed as
$\mathbf{H}_{\text{B},s}^{\text{LoS}}=\boldsymbol{a}_r\left(\theta_{s},\varphi_{s}\right)\boldsymbol{a}_t^{T}\left(\theta_{\text{B}},\varphi_{\text{B}}\right)$ and
$\mathbf{H}_{\text{B},s}^{\text{NLoS}}=\sum\nolimits_{l=1}^{{L}_{\text{B},s}}{{c}_{\text{B},s}^l}{{\mathbf{a}}_{r}}\left( \theta_s^{l},\varphi_s^l \right)\mathbf{a}_{t}^{T}\left(\theta_\text{B}^{l},\varphi_\text{B}^{l} \right)$,
where\par
\begin{footnotesize}
\begin{equation}\label{RSU1}
\begin{aligned}
& {{\mathbf{a}}_{t}}\left( {{\theta }_{\text{B}}},{{\varphi }_{\text{B}}} \right)=\frac{1}{\sqrt{{{N}_{a}}}}[1,\cdots ,{{e}^{j\frac{2\pi }{\lambda }d\left( {{i}_{x,\text{B}}}\cos {{\varphi }_{\text{B}}}\cos {{\theta }_{\text{B}}}+{{i}_{z,\text{B}}}\sin {{\theta }_{\text{B}}} \right)}}, \\
& \cdots ,{{e}^{j\frac{2\pi }{\lambda }d\left( \left( {{L}_{x,\text{B}}}-1 \right)\cos {{\varphi }_{\text{B}}}\cos {{\theta }_{\text{B}}}+\left( {{L}_{z,\text{B}}}-1 \right)\sin {{\theta }_{\text{B}}} \right)}}{{]}^{T}},
\end{aligned}
\end{equation}
\end{footnotesize}%
\begin{footnotesize}
\begin{equation}\label{RSU2}
\begin{aligned}
&{{\mathbf{a}}_{r}}\left( {{\theta }_s},{{\varphi }_s} \right)=\frac{1}{\sqrt{{{N}_{I}}}}[1,\cdots ,{{e}^{j\frac{2\pi }{\lambda }d\left( {{i}_{x,s}}\cos {{\varphi }_s}\cos {{\theta }_s}+{{i}_{z,s}}\sin {{\theta }_s} \right)}}, \\
& \cdots ,{{e}^{j\frac{2\pi }{\lambda }d\left( \left( {{L}_{x,s}}-1 \right)\cos {{\varphi }_s}\cos {{\theta }_s}+\left( {{L}_{z,s}}-1 \right)\sin {{\theta }_s} \right)}}{{]}^{T}},
\end{aligned}
\end{equation}
\end{footnotesize}%
are the BS transmit and the $s$th RIS receive steering vectors, respectively.
$i_x$ and $i_z$ are the antenna indices in the x-z plane,
$d$ and $\lambda$ are the inter-antenna spacing and the signal wavelength.
For LoS component, ${\theta }_{\text{B}}, {\theta }_s \in \left[ -\frac{\pi }{2},\frac{\pi }{2} \right]$ are the elevation AOD and AOA,
and $\varphi_{\text{B}}, \varphi_s \in \left[ 0,{2}{\pi } \right]$ are the azimuth AOD and AOA of the $s$th BS-RIS link, respectively.
For scattering component, $\theta _{\text{B}}^{l}$ and $\theta _s^{l}$ are the elevation AOD and AOA,
and $\varphi _{\text{B}}^{l}$ and $\varphi _s^{l}$ are the azimuth AOD and AOA associated to the $(s, l)$th path, respectively.
${L}_{\text{B},s}$ is the number of paths and ${{c}_{\text{B},s}^{l}}\sim \mathcal{CN}\left( 0,1 \right)$ is the corresponding path gain.
The channel gain between the BS and the $k$th SCU ${{\mathbf{H}}_{\text{B,}k}}\in\mathbb{C}^{{N}_{r}\times{N}_{a}}$ is considered as
\begin{equation}\label{RSUCVU}
{{\mathbf{H}}_{\text{B,}k}}= \sqrt{\frac{{{\lambda }_{\text{B},k}}}{1+{{\lambda }_{\text{B},k}}}}\mathbf{H}_{\text{B},k}^{\text{LoS}}+\sqrt{\frac{1}{1+{{\lambda }_{\text{B},k}}}}\mathbf{H}_{\text{B},k}^{\text{NLoS}} ,
\end{equation}
where ${{\lambda }_{\text{B},k}}$ is the Rician factor of the $k$th BS-SCU link.
$\mathbf{H}_{\text{B},k}^{\text{LoS}}=\boldsymbol{a}_r\left(\hat{\theta}_k,\hat{\varphi}_k\right)\boldsymbol{a}_t^{T}\left(\theta_k,\varphi_k\right)$ and
$\mathbf{H}_{\text{B},k}^{\text{NLoS}}=\sum\nolimits_{l=1}^{{{L}_{\text{B},k}}}{{c}_{\text{B},k}^l}{{\mathbf{a}}_{r}}\left( \hat{\theta}^{l}_{k},\hat{\varphi}^{l}_{k} \right)\mathbf{a}_{t}^{T}\left(\theta^{l}_{k},\varphi^{l}_{k} \right)$ are the LoS and scattering components, respectively.
$\boldsymbol{a}_t\left(\theta_k,\varphi_k\right)\in\mathbb{C}^{{N}_{a}\times1}$ and $\boldsymbol{a}_r\left(\hat{\theta}_k,\hat{\varphi}_k\right)\in\mathbb{C}^{{N}_{r}\times1}$ are the BS transmit and the $k$th SCU's receive steering vectors, respectively.
${{L}_{\text{B},k}}$ is the number of paths associated to SCU $k$ and ${{c}_{\text{B},k}^{l}}\sim \mathcal{CN}\left( 0,1 \right)$ is the $\left( k,l \right)$th path gain.
The channel gain between the $s$th RIS and the $k$th SCU ${\mathbf{H}_{s,k}}\in \mathbb{C}^{{N}_{r}\times{N}_{I}}$ is represented as
\begin{equation}\label{RISSCU}
{{\mathbf{H}}_{s,k}}= \sqrt{\frac{{{\lambda }_{s,k}}}{1+{{\lambda }_{s,k}}}}\mathbf{H}_{s,k}^{\text{LoS}}+\sqrt{\frac{1}{1+{{\lambda }_{s,k}}}}\mathbf{H}_{s,k}^{\text{NLoS}},
\end{equation}
where ${{\lambda }_{s,k}}$ is the Rician factor of the $(s,k)$th RIS-SCU link.
$\mathbf{H}_{s,k}^{\text{LoS}}$ and $\mathbf{H}_{s,k}^{\text{NLoS}}$ are the corresponding LoS and scattering components, respectively.
Likewise, we establish the channel gains between the BS and the Eve and between the $s$th RIS and the Eve, which are defined by $\mathbf{h}_{\text{B},e}\in {{\mathbb{C}}^{N_a}}$ and $\mathbf{h}_{s,e}\in {{\mathbb{C}}^{{{N}_{I}}}}$, respectively.

We adopt symbol-level precoding to perform both radar sensing and multi-user communications.
Specifically, let $\mathbf{x}[n]={{\left[ {x}_{1}[n],...,{x}_{N_a}[n] \right]}^{T}}$ be the precoded vector in the $n$th slot (symbol duration), where $x_m[n]$ denotes the baseband signal sent from the $m$th antenna, $m=1,...,N_a$.
In the sensing context, $\mathbf{x}[n]$ can be considered to be the $n$th sample of the sensing waveform.
Let $\mathbf{s}[n]={{[{{s}_{1}}[n],...,{{s}_{{{N}_{a}}}}[n]]}^{T}}$ be the information symbols sent to the $K$ SCUs.
The technique adopts a non-linear mapping from $\mathbf{s}[n]$ to $\mathbf{x}[n]$, and designs $\mathbf{x}[n]$ directly based on the symbol vector $\mathbf{s}[n]$\footnote{Symbol-level precoding technique can provide more DoFs to enhance both target sensing and multi-user communication capability, as well as ensure superior instantaneous sensing capability.}.
Then, the received baseband signal at the $k$th SCU in the $n$th slot is given by
\begin{equation}\label{SINRRi}
\begin{aligned}
&{\mathbf{r}_{k}}\left( n \right)=\left( \mathbf{H}_{\text{B},k}+\sum\nolimits_{s=1}^{S}\mathbf{H}_{\text{s},k}\mathbf{\Theta}_s\mathbf{G}_s \right)\mathbf{x}\left( n \right)  \\
&~~~~~~~~~~~+\sum\nolimits_{s=1}^{S}{\mathbf{H}_{\text{s},k}\mathbf{\Psi }_\text{s}{{\mathbf{z}}_{0}}}\left( n \right) +{\boldsymbol{\upsilon}_{k}}\left( n \right), \forall k,
\end{aligned}
\end{equation}
where ${{\mathbf{z}}_{0}}(n)\sim \mathcal{CN}(\mathbf{0},\sigma _\text{RIS}^{2}{{\mathbf{I}}_{{{N}_{I}}}})$ and ${\boldsymbol{\upsilon }_{k}}\left( n \right)\sim \mathcal{CN}\left( \mathbf{0},\sigma _{\text{C},k}^{2}{\mathbf{I}}_{{N}_{r}} \right)$ are the additive white Gaussian noise (AWGN) at the active reflection units and the $k$th SCU, respectively.
By applying a receive beamformer $\mathbf{v}_{k}\in\mathbb{C}^{N_{r}}, \forall k,$ on ${\mathbf{r}_{k}}\left( n \right)$ to recover the data signals of the $k$th SCU, we have
\begin{equation}
\begin{aligned}
&{y}_{k}\left( n \right)=\mathbf{v}_k^{H}\mathbf{H}_{k}\mathbf{x}\left( n \right)+\sum\nolimits_{s=1}^{S}\mathbf{v}_k^{H}{\mathbf{H}_{\text{s},k}\mathbf{\Psi }_{\text{s}}{{\mathbf{z}}_{0}}}\left( n \right) \\
&~~~~~~~~~~+\mathbf{v}_k^{H}{\boldsymbol{\upsilon}_{k}}\left( n \right), \forall k,
\end{aligned}
\end{equation}
where $\mathbf{H}_{k}=\mathbf{H}_{\text{B},k}+\sum\nolimits_{s=1}^{S}\mathbf{H}_{\text{s},k}\mathbf{\Theta}_{\text{s}}\mathbf{G}_{\text{s}}$.
Similarly, the received signal at the Eve in the $n$th slot is expressed as
\begin{equation}\label{SINRRi1}
\begin{aligned}
&{{y}_{e}}\left( n \right)=\left( \mathbf{h}_{\text{B},e}^{H}+\sum\nolimits_{s=1}^{S}\mathbf{h}_{\text{s},e}^{H}\mathbf{\Theta}_\text{s}\mathbf{G}_\text{s}\right)\mathbf{x}\left( n \right) \\
&~~~~~~~~~~+\sum\nolimits_{s=1}^{S}{\mathbf{h}_{\text{s},e}^{H}\mathbf{\Psi}_\text{s}{{\mathbf{z}}_{0}}}\left( n \right)+{{\upsilon}_{e}}\left( n \right),
\end{aligned}
\end{equation}
where ${{\upsilon }_{e}}\left( n \right)\sim \mathcal{CN}\left( 0,\sigma _{\text{e}}^{2} \right)$ represents the AWGN at the Eve.}
%As discussed in~\cite{2016Kalantari}, the investigation of the DM scheme is based on strict phase and relaxed phase requirements.
%As to the strict phase signal optimization, the signal received at the $k$th SCU ${{y}_{k}}\left( n \right)$ must have the identical phase as the desired information symbol (namely, $s_{k}\left( n \right)$).
%This requirement seriously limits the DoFs in optimizing the transmit signal.
%Thus, based on the idea of CI~\cite{2015Masouros}, the communication quality constraint is considered to locate the desired information symbol for all SCUs within a constructive area instead of constraining the intended symbol in the vicinity of the constellation point, i.e.,
Based on the concept of CI~\cite{2015Masouros}, the communication quality-of-service (QoS) constraint is considered to locate the received symbol for each SCU within a constructive area instead of constraining the intended symbol in the proximity of the constellation point, i.e.,
\begin{equation}\label{SNRCt}
\left| \arg \left( \mathbf{v}_k^{H}\mathbf{H}_{k}\mathbf{x}\left( n \right) \right)-\arg \left( s_{k}\left( n \right) \right) \right|\le \delta, \forall k,
\end{equation}
where $\delta$ denotes the phase threshold where the interference-excluding received symbols are designed to lie.
Since the CI-type waveform design aims to convert the inherent MUI into useful power by pushing the received signal away from the valid decision regions, undesired interference can be exploited and becomes the useful received power.
Taking the quadrature phase shift keying (QPSK) modulation as an example, the constructive area is denoted as the white zone that benefits the signal demodulation at the SCUs, as presented in Fig.~\ref{CI}.
Since $\left| s_{k}\left( n \right) \right|=1$, the interference-excluding signal ${{\bar{y}}_{k}}\left( n \right)$ is given by
\begin{equation}\label{baryk}
{{{\bar{y}}}_{k}}\left( n \right)=\mathbf{v}_k^{H}\mathbf{H}_{k}\mathbf{x}\left( n \right)\frac{s_{k}^{*}\left( n \right)}{\left| s_{k}\left( n \right) \right|}
 =\mathbf{v}_k^{H}\bar{\mathbf{H}}_{k}\mathbf{x}\left( n \right),  \forall k,
\end{equation}
where $\bar{\mathbf{H}}_{k}=\mathbf{H}_{k}s_{k}^{*}$.
Then, the CI-type communication QoS constraint is reformulated as~\eqref{barhkn}, where $\phi_k =\pm \pi /M$,
$\mathbf{\Lambda }_k=\sum\nolimits_{s=1}^{S}\mathbf{H}_{\text{s},k}^{H}\mathbf{\Psi }_s\mathbf{\Psi}_s^H\mathbf{H}_{\text{s},k}$
and ${{\Gamma }_{k}}$ is the certain SINR threshold for the $k$th SCU.
\begin{figure*}
\begin{equation}\label{barhkn}
\left| \operatorname{Im}\left(\mathbf{v}_k^{H}\bar{\mathbf{H}}_{k}\mathbf{x}\left( n \right) \right) \right|\le  \left( \operatorname{Re}\left(\mathbf{v}_k^{H}\bar{\mathbf{H}}_{k}\mathbf{x}\left( n \right) \right)-\sqrt{{\Gamma }_{k}\left(\sigma _\text{RIS}^{2}\mathbf{v}_k^H\mathbf{\Lambda }_k\mathbf{v}_k+\sigma _{k}^2\right)}\right)\tan {{\phi }_{k}},\forall k,n,
\end{equation}
\hrulefill
\end{figure*}
\begin{figure}[ht]
\captionsetup{font={footnotesize}}
	\begin{center}
	{\includegraphics[width=5.5cm]{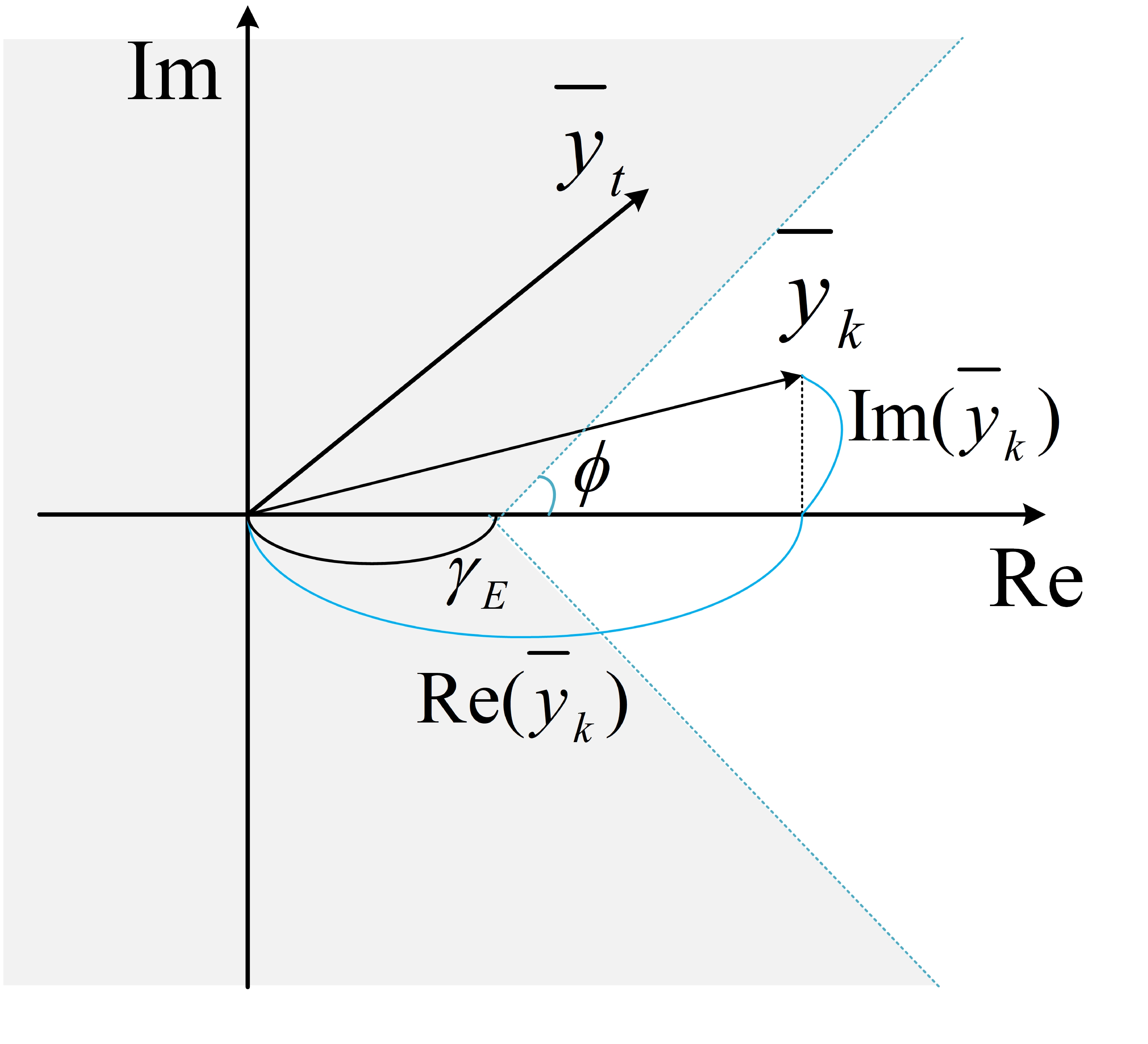}}
		\caption{Depiction of the CI using QPSK as an example, where the white and gray zones denote the constructive and destructive areas.}
		\label{CI}
	\end{center}
\end{figure}
While the CI-type precoding method can ensure satisfactory low symbol error rate (SER) at each SCU, we also consider the detection capability at the Eve to avoid the confidential information from being decoded.
To deceive the Eve, we propose to manipulate the confidential signal eavesdropped at the Eve by imposing the signal to lie in the destructive zone for detection, presented as the gray zone in Fig.~\ref{CI}.
Hence, the signals received at the Eve are pushed away from the valid decision zone such that precision detection at the Eve becomes challenging or even impossible.
Similarly, the received interference-less signal at the Eve is given by
\begin{equation}\label{baryt}
\begin{aligned}
&{{{\bar{y}}}_{e}}\left( n \right)=\left( \mathbf{h}_{\text{B},e}^{H}+\sum\nolimits_{s=1}^{S}\mathbf{h}_{\text{s},e}^{H}\mathbf{\Theta}_\text{s}\mathbf{G}_\text{s}\right)  \mathbf{x}\left( n \right) \frac{s_{k}^{*}\left( n \right)}{\left| {{s}_{k}}\left( n \right) \right|}\\
&~~~~~~~~=\bar{\mathbf{h}}_{e}^{H}\mathbf{x}\left( n \right),
\end{aligned}
\end{equation}
where $\bar{\mathbf{h}}^{H}_{e}=\left( \mathbf{h}_{\text{B},e}^{H}+\sum\nolimits_{s=1}^{S}\mathbf{h}_{\text{s},e}^{H}\mathbf{\Theta}_\text{s}\mathbf{G}_\text{s}  \right) s_{k}^{*}$.
Hence, the DI-type communication security constraint is given by~\eqref{barhen1}, where ${{\Gamma }_{E}}$ corresponds to the SINR related scalar $\gamma_E$, which is the threshold to the decision zone of the received information symbol at Eve.
\begin{figure*}
\begin{equation}\label{barhen1}
\left| \operatorname{Im}\left( \mathbf{\bar{h}}_{e}^{H}(n)\mathbf{x}\left( n \right)  \right) \right|\ge
\left( \operatorname{Re}\left( \mathbf{\bar{h}}_{e}^{H}\left(n\right)\mathbf{x}\left(n\right) \right)-\sqrt{{\Gamma }_{\text{E}}\left(\sigma _\text{RIS}^{2}\sum\nolimits_{s=1}^{S}\mathbf{h}_{\text{s},e}^{H}\mathbf{\Psi }_\text{s}\mathbf{\Psi }_\text{s}^H\mathbf{h}_{\text{s},e}^{H}+\sigma _{e}^2\right)}\right) \tan {{\phi }_{\text{e}}},  \forall n,
\end{equation}
\end{figure*}
%In this case, we express~\eqref{barhen1} as the DI security constraint.
%Moreover, we assume that the CSI of the Eve is exactly known at the BS's receiver.
%According to these assumptions, our research can offer the best security performance bounds for actual hybrid RIS-enhanced ISAC application scenarios.
%Concerning a high resolution sensing system, the probed target is usually regarded as a set of multiple scattering points.
%Moreover, the extended target is characterized by the target geometry property and orientation with respect to the LoS.  %~\cite{2000Jacobs}
%Fig.~\ref{extended} shows how the TIR amplitude is modeled employing the projection of the multiple scattering points on the LoS.
%Thus, considering TAA coefficient uncertainty signifies the critical issue for an effective hybrid RIS-enhanced ISAC system optimization.

\subsection{Sensing Model}

To perform target detection, the BS radiates the ISAC probing signal towards the eavesdropping target and observes its return signals reflected by the target in the vicinity concurrently.
The effective channels from the BS and RIS to the target are LoS and determined by its location.
{\color{blue}Generally, the TL is imperfectly known to the RIS-aided ISAC system owing to its movement and random fluctuation.
Thus, the target is assumed to locate in an uncertain angular interval $\bar{\chi }=[\theta_0 -\Delta \theta ,\theta_0 +\Delta \theta ]$ and $\hat{\chi }=[\varphi_0 -\Delta \varphi ,\varphi_0+\Delta \varphi ]$.
Let $\{\theta^{p},\varphi^{p}\}\in \text{card}\left( {\bar{\chi }}, {\hat{\chi }} \right)$ denotes the $p$th possible location in the certain region, where $\text{card}\left( \cdot \right)$ denotes the cardinality of $\left( \cdot \right)$.}
The effective channels for the target return and the clutter return are defined respectively as\par
\begin{footnotesize}
\begin{subequations}\label{Htc}
\begin{align}
&\mathbf{H}_{\mathrm{e}}^p\!=\!
\left(\mathbf{a}_{\mathrm{B},e}^p\!+\!\sum\limits_{s=1}^{S}\mathbf{G}_s^H\mathbf{\Theta}_s^H\mathbf{a}_{s,e}^p\right)
\left(\mathbf{a}_{\mathrm{B},e}^{pH}\!+\!\sum\limits_{s=1}^{S}\mathbf{a}_{s,e}^{pH}\mathbf{\Theta}_s\mathbf{G}_s\right), \forall p, \\
&\mathbf{H}_{\mathrm{c}}\!=\!
\left(\mathbf{a}_{\mathrm{B},c}\!+\!\sum\limits_{s=1}^{S}\mathbf{G}_s^H\mathbf{\Theta}_s^H\mathbf{a}_{s,c}\right)
\left(\mathbf{a}_{\mathrm{B},c}^H\!+\!\sum\limits_{s=1}^{S}\mathbf{a}_{s,c}^H\mathbf{\Theta}_s\mathbf{G}_s\right),
\end{align}
\end{subequations}
\end{footnotesize}%
where
$\mathbf{a}_{\text{B},e}^p=\sqrt{{{l}_{\text{B},e}}}\mathbf{a}\left( {{\theta }_{\text{B},e}^p},{{\varphi}_{\text{B},e}^p} \right)$ and
$\mathbf{a}_{\text{B},c}=\sqrt{{{l}_{\text{B},c}}}{{\mathbf{a}}}\left( {{\theta }_{\text{B},c}}, {{\varphi }_{\text{B},c}} \right)$ are the steering vectors from the BS to the $p$th TL and interference, respectively.
$\mathbf{a}_{s,e}^p=\sqrt{{{l}_{s,e}}}\mathbf{a}\left( {{\theta }_{s,e}^p},{{\varphi}_{s,e}^p} \right)$ and
$\mathbf{a}_{s,c}=\sqrt{{{l}_{s,c}}}\mathbf{a}\left( {{\theta }_{s,c}^p},{{\varphi}_{s,c}^p} \right)$ are the steering vectors from the $s$th RIS to the $p$th TL and interference, respectively.
${{l}_{s,e}}$, ${{l}_{s,c}}$, ${{l}_{\text{B},e}}$, and ${{l}_{\text{B},c}}$ are the path loss from the $s$th RIS/BS to the target and interference, respectively.
${{{\theta }_{s,e}}}$, ${{\theta }_{s,c}}$, ${{{\theta }_{\text{B},e}}}$, and ${{\theta }_{\text{B},c}}$ are the elevation AOD from the $s$th RIS/BS to the target and interference, respectively.
${{{\varphi }_{s,e}}}$, ${{\varphi }_{s,c}}$, ${{{\varphi }_{\text{B},e}}}$, and ${{\varphi }_{\text{B},c}}$ are the azimuth AOD from the $s$th RIS/BS to the target and interference, respectively.

For a sensing system using high range resolution, the probed target should be modeled as a set of multiple scattering points.
The scattering characteristic of the extended target is specified by the target impulse response (TIR).
The TIR relies on the target orientation~\cite{2011Shi}, i.e., the scattering characteristic is sensitive with respect to (w.r.t.) a fluctuation of the target aspect angle (TAA).
For a specific range cell $n$ and TAA $\hat \theta$, the TIR matrix is given by ${{\mathbf{T}}_{\hat \theta }}\left( n \right)\in {{\mathbb{C}}^{{{N}_{a}}\times {{N}_{a}}}}$\footnote{$\mathbf{T}_{\hat \theta }^{k,l}\left( n \right)$ denotes TIR coefficient between the $l$th transmit antenna and the $k$th receive antenna.}.
The support of the TIR has size $Q$ and the observations are gathered within an interval $M$, with $M=Q+N-1$.
%The coefficient $Q$ is known, since it is dependent on the maximum target size as well as system resolution, namely, ${{\mathbf{T}}_{\theta }}\left( n \right)\approx 0$ unless $n\in \left\{ 1,...,Q \right\}$.
Analogously, the clutter scattering matrix associated with the $n$th range cell is given by $\mathbf{C}\left( n \right)\in {{\mathbb{C}}^{{{N}_{a}}\times {{N}_{a}}}}$.
{\color{blue}Therefore, the radar echo at BS in the $n$th slot is represented as~\eqref{rn}, where ${{\hat{\mathbf{H}}}_\text{bre}}^p=\sum\nolimits_{s=1}^{S}\mathbf{a}_{\text{B},e}^p\mathbf{a}_{s,e}^{pH}\mathbf{\Psi }_s$.
$v_0$ and $v_1$ are the normalized Doppler frequency shifts of the target and the clutter term in radians, respectively.}
\begin{figure*}
\begin{equation}\label{rn}
\mathbf{r}\left( n \right)={{\mathbf{H}}^p_\text{e}}\left( \sum\nolimits_{k=1}^{N}{{\mathbf{T}}_{\hat \theta }}\left( n-k \right)\mathbf{x}\left( k \right)e^{j2\pi(k-1)\nu_0} \right)
+{{\mathbf{H}}_\text{c}}\left( \sum\nolimits_{k=1}^{N}{\mathbf{C}\left( n-k \right) \mathbf{x}\left( k \right)} e^{j2\pi(k-1)\nu_1}\right)
+{{\mathbf{\hat H}}_\text{bre}^p}{{\mathbf{z}}_{1}}\left( n \right)+\boldsymbol{\upsilon }\left( n \right),
\end{equation}
\hrulefill
\end{figure*}
$\mathbf{z}_1(n)\sim \mathcal{CN}\left( \mathbf{0},\sigma _\text{RIS}^{2}{{\mathbf{I}}_{{{N}_{I}}}} \right)$ and $\boldsymbol{\upsilon }(n)\sim
\mathcal{CN}\left( \mathbf{0},\sigma _{R}^{2}{{\mathbf{I}}_{{{N}_{a}}}} \right)$ are additive noise components at the active reflection units and the BS, respectively\footnote{Since the noise component $\mathbf{z}_0(n)$ experiences multiple attenuations via the RIS-Eve-BS and the RIS-Eve-RIS-BS channels when the return signals reach the BS, their power is much less than that of other echo signals, thus can be neglected~\cite{HybridAP1,2023Nguyen}.}.

Collecting all the observations into a single vector by defining $\mathbf{r}= {{\left[ {{\mathbf{r}}^{T}}\left( 1 \right),\ldots,{{\mathbf{r}}^{T}}\left( M \right) \right]}^{T}}\in {{\mathbb{C}}^{{{N}_{a}}M}}$, $\mathbf{x}={{\left[ {{\mathbf{x}}^{T}}\left( 1 \right),\ldots,{{\mathbf{x}}^{T}}\left( N \right) \right]}^{T}}\in {{\mathbb{C}}^{{{N}_{a}}N}}$,
$\mathbf{z }_1={{\left[ {\mathbf{z }_1^{T}}\left( 1 \right),\ldots,{\mathbf{z }_1^{T}}\left( M \right) \right]}^{T}}\in {{\mathbb{C}}^{{{N}_{I}}M}}$,
$\mathbf{p}(\nu_0) = \begin{bmatrix}
		1, e^{j2\pi\nu_0}, \ldots, e^{j2\pi(N-1)\nu_0}
	\end{bmatrix}^T$,
$\mathbf{p}(\nu_1) = \begin{bmatrix}
		1, e^{j2\pi\nu_1}, \ldots, e^{j2\pi(N-1)\nu_1}
	\end{bmatrix}^T$
and $\boldsymbol{\upsilon }={{\left[ {{\boldsymbol{\upsilon }}^{T}}\left( 1 \right),\ldots,{{\boldsymbol{\upsilon }}^{T}}\left( M \right) \right]}^{T}}\in {{\mathbb{C}}^{{{N}_{a}}M}}$,
the radar echo at BS over the $M$ slot is rewritten as
\begin{equation}\label{r2}
\begin{aligned}
 \mathbf{r}={{\mathbf{\bar{H}}}^p_{\text{e}}}\mathbf{\bar{T}}\left( {{ \hat \theta }} \right)\bar{\mathbf{P}}(\nu_0) \mathbf{x}  +{{\mathbf{\bar{H}}}_\text{c}}\mathbf{\bar{C}}\bar{\mathbf{P}}(\nu_1)\mathbf{x}
+{{\mathbf{\tilde{H}}}^p_\text{e}}{{\mathbf{z}}_{1}}+\boldsymbol{\upsilon },
\end{aligned}
\end{equation}
where ${{\mathbf{\bar{H}}}^p_{\text{e}}}= \mathbf{I}_{M}\otimes {{\mathbf{H}}^p_{\text{e}}}$,
${{\mathbf{\bar{H}}}_\text{c}}=\mathbf{I}_{M}\otimes {{\mathbf{H}}_{c}}$,
${{\mathbf{\tilde{H}}}_\text{e}^p}= {{\mathbf{I}}_{M}}\otimes {\hat{\mathbf{H}}_\text{bre}^p}$,
$\bar{\mathbf{P}}(\nu_0)= \mathbf{I}_{N_a}\otimes \text{diag} \{ \mathbf{p}(\nu_0) \}$,
$\bar{\mathbf{P}}(\nu_1)= \mathbf{I}_{N_a}\otimes  \text{diag} \{\mathbf{p}(\nu_1)\}$,
$\mathbf{\bar{T}}\left(\hat \theta  \right)=\sum\limits_{n=1}^{Q}{{{\mathbf{J}}_{n}}}\otimes {{\mathbf{T}}_{ \hat \theta }}(n)$ and
$\mathbf{\bar{C}}=\sum\limits_{n=-N+1}^{M-1}{{{\mathbf{J}}_{n}}}\otimes \mathbf{C}\left( n \right)$ are the $N_aM\times N_aN$ dimensional TIR and clutter impulse response (CIR) matrices, respectively,
with ${{\mathbf{J}}_{n}}$ the shift matrix\footnote{The shift matrix with delay $n$ is $${{\mathbf{J}}_{n}}({{l}_{1}},{{l}_{2}})=\left\{ \begin{matrix}
				1,\text{    if }{{l}_{1}}-{{l}_{2}}=n  \\
				0,\text{    if }{{l}_{1}}-{{l}_{2}}\ne n  \\
			\end{matrix} \right.,
{{l}_{1}}\in \left\{ 1,\cdots ,M \right\},{{l}_{2}}\in \left\{ 1,\cdots ,N \right\}.$$}.
For each clutter scatterer in the $n$th range cell, we assume that it is zero mean, i.e., $\mathbb{E}\left[ \mathbf{C}\left( n \right) \right]=\mathbf{0}$, and define ${{r}_{kl}}\left( n,{n}' \right)=\mathbb{E}\left[ {C}_{k}\left( n \right){C}_{l}^{*}\left( {{n}'} \right) \right]$ as the correlation coefficient between the clutter scatterers of the $j$ link and the $l$ link at $n$th and ${n}'$th range cells.

%We focus on a radar target that has been tracked, and consider that its TAA is imperfectly available according to the previous observations and prescan procedures~\cite{2022YangJing}.
%This means that the exact TAA is not available, resulting in the uncertainty on the target scattering matrix.
%\footnote{The parameter $I$ is set according to the following rule: $I\Delta \theta ={{\theta }_{2}}-{{\theta }_{1}}$.} each regulated to a given TAA and target location ${{\theta }_{i}},i=1,...,I$.
%Specifically, the likely TAA belongs to a uncertainty set, i.e., $I_{{{\hat \theta }_{1}}}^{{{\hat \theta }_{2}}}=\left\{ \left. i\Delta \hat \theta  \right|i\in \mathbb{Z},i\Delta \hat \theta \in \left[ {{\hat \theta }_{1}},{{\hat \theta }_{2}} \right] \right\}$, where $\Delta \hat \theta$ is the sampling step.
%we propose an array of space-time linear filters, each turned to a given TAA and target location, to design the considered system.
{\color{blue}We consider the scenario where a rough estimation of the TL is available at the BS and RISs.
Thus, to evade the uncertainty on the CSI of eavesdropping target, we propose an array of space-time linear receive filters to design the hybrid RIS-aided ISAC systems.}
Exactly, assuming that the observations $\mathbf{r}$ is processed through the bank of filters $\{\mathbf{w}_p\}_{p=1}^P \in {{\mathbb{C}}^{{{N}_{a}}M}}$ each turned to a given TL $\{\theta_p, \varphi_p\}, \forall p$.
Then, the $p$th radar output is expressed as
\begin{equation}\label{barH}
\begin{aligned}
&{{y}}_p=\mathbf{w}^{H}_p\mathbf{r}=\mathbf{w}^{H}_p{{\mathbf{\bar{H}}}^p_\text{e}}\mathbf{\bar{T}}\left( {{ \hat \theta }} \right)\bar{\mathbf{P}}(\nu_0)\mathbf{x}
  \\
&+\mathbf{w}^{H}_p{{\mathbf{\bar{H}}}_{\text{c}}}\mathbf{\bar{C}}\bar{\mathbf{P}}(\nu_1)\mathbf{x}
+\mathbf{w}^{H}_p{{\mathbf{\tilde{H}}}_{e}^p}{{\mathbf{z}}_{1}}+\mathbf{w}^{H}_p\boldsymbol{\upsilon }. \\
\end{aligned}
\end{equation}
Hence, the corresponding received SINR is denoted as
\begin{equation}\label{SINRRi}
\text{SINR}_{p}=\frac{{{\left| \mathbf{w}^{H}_p{{\mathbf{A}}_{p}}\mathbf{x} \right|}^{2}}}
{\mathbf{w}^{H}_p\left( {{\mathbf{\Pi }}_{c}}+\mathbf{\Sigma }_p+\mathbf{I}\sigma _{R}^{2} \right)\mathbf{w}_p},
\end{equation}
where ${\mathbf{A}}_{p}={{\mathbf{\bar{H}}}^p_\text{e}}\mathbf{\bar{T}}\left( {{\hat \theta }} \right)\bar{\mathbf{P}}(\nu_0)$, ${{\mathbf{\Pi}}_{c}}={{\mathbf{\bar{H}}}_{\text{c}}}\mathbf{\bar{C}}\bar{\mathbf{P}}(\nu_1)\mathbf{x}\mathbf{x}^H\bar{\mathbf{P}}^H(\nu_1)\mathbf{\bar{C}}^H{{\mathbf{\bar{H}}}_{\text{c}}^H}$, and $\mathbf{\Sigma }_p=\sigma _{\text{RIS}}^{2}{{\mathbf{\tilde{H}}}_{e}^p}{{\mathbf{\tilde{H}}}_{e}^{pH}}$.
The power required by the active RIS is
\begin{equation}\label{PR}
{{P}_{\text{R},s}}=\mathbf{x}^{H}\mathbf{U}\mathbf{x}+\sigma _{\text{RIS}}^{2}\text{tr}\left( {{\mathbf{\Psi }}^{H}_s}\mathbf{\Psi }_s \right),  \forall s,
\end{equation}
where $\mathbf{U}={{\left( \mathbf{e}_{n}^{T}\otimes \mathbf{G}_s \right)}^{H}}{{\mathbf{\Psi }}^{H}_s}\mathbf{\Psi }_s\left( \mathbf{e}_{n}^{T}\otimes \mathbf{G}_s \right)$.
We remark that for a specified probability of false alarm, the target detection performance monotonically improves with the received SINR under the additive Gaussian noise~\cite{2015MaioDe}.

\subsection{Problem Formulation}\label{Formulation}

%As the TAA is uncertain, the joint design of the transceiver of ISAC and the weights of the hybrid RIS should consider the synchronous optimization of multiple objectives, that is $\text{SINR}_{i}^{\text{R}}\left( \mathbf{x},\boldsymbol{\vartheta},\left\{ \mathbf{w}_{i} \right\}_{i=1}^{I} \right),i=1,...,I$.
%This method relies on objectives prioritization and focuses on a specific weighted sum of objectives to assess system capability; evidently, the average performance metric is equivalent to handle multi-objective Pareto optimization exploiting scalarization.
%Considering imperfect knowledge of the Eve's TAA and CSI,

{\color{blue}Our goal is to jointly design the transmit waveform and the receive filters at the BS $\mathbf{x}$, $\{\mathbf{w}_{p}\}^P_{p=1}$, the weights at the hybrid RIS $\mathbf{\Theta}$,
and the SCUs' receive beamformers $\{\mathbf{v}_{k}\}^K_{k=1}$ for the considered system.
We consider maximizing the worst-case sensing SINR among all possible TLs while ensuring the CI-type communication QoS and DI-type security requirements, the RIS weight constraint, and total power budget constraint.
Hence, the optimization problem can be formulated as
\begin{subequations}\label{P0}
\begin{align}
&\underset{  \mathbf{x},\mathbf{w}_p, \mathbf{\Theta},{\mathbf{v}}_{k}}{\mathop{\max }}\min_{\theta^p,{\varphi}^p}\text{SINR}_{p} \label{P01}  \\
&~~~~~~~~~\text{s}\text{.t}\text{. }~{{\left\| \mathbf{x} \right\|}^{2}}\le {{P}_{\max}^{\text{B}}}, \label{P02}  \\
&~~~~~~~~~~~~~~{{P}_{\text{R},s}}\le P_{\max }^{\text{R}}, \forall s,   \label{P03} \\
&~~~~~~~~~~~~~~{{e}^{j\frac{2\pi n}{{{2}^{b}}}}},n\in \left\{ 0,1,...,{{2}^{b}}-1 \right\},n\in {{N}_{I}},    \label{P04} \\
&~~~~~~~~~~~~~~ \eqref{barhkn},~\eqref{barhen1}, \label{P05}
\end{align}
\end{subequations}
where ${{P}_{\max}^{\text{B}}}$ and $ P_{\max }^{\text{R}}$ are the maximum power budget of BS and hybrid RIS, respectively.
\eqref{P04} is the discrete reflection coefficient (DRC) constraints.
Problem~\eqref{P0} poses several challenges, mainly owing to the presence of the quartic objective function~\eqref{P01}.
Moreover, the highly coupled optimization variables further aggravate the challenge.
To address this challenge, we split the joint design problem into four subproblems, and solve effectively them via GFP, PDD and PCCP methods in the following section.}

\section{Hybrid RIS-Enhanced ISAC Optimization}\label{ProbForm}

\subsection{Solve for $\{\mathbf{w}_{p}\}^P_{p=1}$ and $\{\mathbf{v}_{k}\}^K_{k=1}$}

In~\eqref{P0}, it is easy to find that the receive filter bank $\{\mathbf{w}_{p}\}^P_{p=1}$ only exists in the objective function and they are independent from each other.
The optimization for $\{\mathbf{w}_{p}\}^P_{p=1}$ is formulated as a minimum variance distortionless response (MVDR) problem, whose closed-form solution is
\begin{equation}\label{wwi}
{\mathbf{w}}^{*}_p=\frac{{{\left( {{\mathbf{\Pi }}_{c}}+\mathbf{\Sigma }_p+\sigma _{R}^{2}\mathbf{I} \right)}^{-1}}{{\mathbf{A}}_{p}}\mathbf{x}}
{{{\left\| {{\left( {{\mathbf{\Pi }}_{c}}+\mathbf{\Sigma }_p+\sigma _{R}^{2}\mathbf{I} \right)}^{-1/2}}{{\mathbf{A}}_{p}}\mathbf{x} \right\|}^{2}}}, \forall p.
\end{equation}
{\color{blue}Fixed other variables, the optimization problem w.r.t. the SCUs' receive beamformers $\{\mathbf{v}_{k}\}^K_{k=1}$ is reduced to a feasibility check problem\par
\begin{footnotesize}
\begin{subequations}\label{30}
\begin{align}
&~~~\text{Find }~~~~~~~~~~~~~~~~~~~~~~~ \mathbf{v}_k \\
&\text{ s.t. } \left( \sqrt{{\Gamma }_{k}\left(\sigma _\text{RIS}^{2}\mathbf{v}_k^H\mathbf{\Lambda}_k\mathbf{v}_k+\sigma _{k}^2\right)}-\operatorname{Re}\left(\mathbf{v}_k^{H}\bar{\mathbf{H}}_{k}\mathbf{x}\left( n \right) \right)\right) \notag \\
&~~~~~~~~~\times\tan {{\phi }_{k}}-\operatorname{Im}\left(\mathbf{v}_k^{H}\bar{\mathbf{H}}_{k}\mathbf{x}\left( n \right) \right)\leq 0, \forall k,n,\\
&~~~~~~\left( \sqrt{{\Gamma }_{k}\left(\sigma _\text{RIS}^{2}\mathbf{v}_k^H\mathbf{\Lambda}_k\mathbf{v}_k+\sigma _{k}^2\right)}-\operatorname{Re}\left(\mathbf{v}_k^{H}\bar{\mathbf{H}}_{k}\mathbf{x}\left( n \right) \right)\right) \notag \\
&~~~~~~~~~\times\tan {{\phi }_{k}}+\operatorname{Im}\left(\mathbf{v}_k^{H}\bar{\mathbf{H}}_{k}\mathbf{x}\left( n \right) \right)\leq 0,  \forall k,n,
\end{align}
\end{subequations}
\end{footnotesize}%
To solve it, we consider another closely related problem as follows\par
\begin{footnotesize}
	\begin{subequations}\label{31}
		\begin{align}
			&~~~\min_{\mathbf{v}_k,c_k}~~~~~~~~~~~~~~~~~~~~~~c_k \\
			&\text{ s.t. } \left( \sqrt{{\Gamma }_{k}\left(\sigma _\text{RIS}^{2}\mathbf{v}_k^H\mathbf{\Lambda}_k\mathbf{v}_k+\sigma _{k}^2\right)}-\operatorname{Re}\left(\mathbf{v}_k^{H}\bar{\mathbf{H}}_{k}\mathbf{x}\left( n \right) \right)\right)  \notag\\
			&~~~~~~~~~\times\tan {{\phi }_{k}}-\operatorname{Im}\left(\mathbf{v}_k^{H}\bar{\mathbf{H}}_{k}\mathbf{x}\left( n \right) \right)\leq c_k, \forall k,n, \label{31a}\\
			&~~~~~~\left( \sqrt{{\Gamma }_{k}\left(\sigma _\text{RIS}^{2}\mathbf{v}_k^H\mathbf{\Lambda}_k\mathbf{v}_k+\sigma _{k}^2\right)}-\operatorname{Re}\left(\mathbf{v}_k^{H}\bar{\mathbf{H}}_{k}\mathbf{x}\left( n \right) \right)\right)  \notag\\
			&~~~~~~~~~\times\tan {{\phi }_{k}}+\operatorname{Im}\left(\mathbf{v}_k^{H}\bar{\mathbf{H}}_{k}\mathbf{x}\left( n \right) \right)\leq c_k,  \forall k,n, \label{31b}
		\end{align}
	\end{subequations}
\end{footnotesize}%
Note that \eqref{31} is assured to have a feasible solution yielding non-positive $c_k$ if the whole iteration starts from a feasible point.
Minimizing~\eqref{31} is meant to find a more feasible $c_k$, which provides a larger margin to satisfy the constraints \eqref{31a},~\eqref{31b} and thus benefits the optimization of other variables.}
Next, we consider the optimization of transmit beamforming $\mathbf{x}$ and hybrid RIS weights $\mathbf{\Theta}$ with fixed $\{\{\mathbf{w}_{p}\}^P_{p=1}, \{\mathbf{v}_{k}\}^K_{k=1}\}$.
Based on the big-M continuous relaxation technique~\cite{2013Cheng}, we introduce binary variables ${{\eta }_{n,p}}\in \left\{ 0,1 \right\},\forall n,p$ and a sufficiently large factor $\Upsilon >0$, the DI security constraint~\eqref{barhen1} is reformulated as~\eqref{barhtn3}.
\begin{figure*}
\begin{subequations}\label{barhtn3}
\begin{align}
&\left( \operatorname{Re}\left( \mathbf{\bar{h}}_{e}^{pH}\left(n\right)\mathbf{x}\left(n\right) \right)-\sqrt{{\Gamma }_{\text{E}}\left(\sigma _\text{RIS}^{2}\sum\limits_{s=1}^{S}\mathbf{h}_{\text{s},e}^{H}\mathbf{\Psi }_\text{s}\mathbf{\Psi }_\text{s}^H\mathbf{h}_{\text{s},e}^{H}+\sigma _{e}^2\right)}\right)\tan {{\phi }_{\text{e}}}-\operatorname{Im}\left( \mathbf{\bar{h}}_{e}^{pH}(n)\mathbf{x}\left( n \right)  \right)-{{\eta }_{n,p}}\Upsilon \le 0,\forall n,p,  \label{barhtn31} \\
 &\left( \operatorname{Re}\left( \mathbf{\bar{h}}_{e}^{pH}\left(n\right)\mathbf{x}\left(n\right) \right)\!-\!\sqrt{{\Gamma }_{\text{E}}\left(\sigma _\text{RIS}^{2}\sum\limits_{s=1}^{S}\mathbf{h}_{\text{s},e}^{H}\mathbf{\Psi }_\text{s}\mathbf{\Psi }_\text{s}^H\mathbf{h}_{\text{s},e}^{H}\!+\!\sigma _{e}^2\right)}\right)\tan {{\phi }_{\text{e}}}\!+\!\operatorname{Im}\left( \mathbf{\bar{h}}_{e}^{pH}(n)\mathbf{x}\left( n \right)  \right)-(1-{{\eta }_{n,p}})\Upsilon \le 0,\forall n,p. \label{barhtn32}
\end{align}
\end{subequations}
\hrulefill
\end{figure*}
Consequently, the original optimization problem~\eqref{P0} can be recast as
\begin{subequations}\label{P1}
\begin{align}
  &\underset{\mathbf{x}, \mathbf{\Theta}, \eta}{\mathop{\max }}\, \min_{\theta^p,{\varphi}^p }\frac{{{\left| \mathbf{w}^{H}_p{\mathbf{A}}_{p}\mathbf{x} \right|}^{2}}}{\mathbf{w}^{H}_p\left( {{\mathbf{\Pi }}_{c}}+\mathbf{\Sigma }_p+\mathbf{I}\sigma _{R}^{2} \right)\mathbf{w}_p} \nonumber \\
 &~~+\sum\nolimits_{p=1}^{P}\omega \left( \sum\nolimits_{n=1}^{N}{{{\eta }_{n,p}^2}}-\sum\nolimits_{n=1}^{N}{\eta _{n,p}} \right)  \label{P11} \\
 &~~~~~ \text{s.t.}~~\eqref{P02},~\eqref{P03},~\eqref{P04},  \label{P12} \\
 &~~~~~~~~~~~\eqref{barhkn},~\eqref{barhtn31},~\eqref{barhtn32}, \color{blue}\text{0}\le {{\eta }_{n,p}}\le 1,\forall n,p, \label{P15}
\end{align}
\end{subequations}
where $\omega$ is a penalty factor for controlling the objective function.
{\color{blue}We prove that problem~\eqref{P1} is solvable and describe an algorithm that finds a high-quality solution.
To this end, we exploit some results from the GFP theory, which are summarized here in the form of a lemma.
\begin{lemma}
Let $\mathcal{X}\in \mathbb{C}^N$ be a convex compact set,
$f_{p}\left(\mathbf{x}\right)$ be non-negative concave functions over $\mathcal{X}$,
and $g_{p}\left(\mathbf{x}\right)$ be positive convex functions over $\mathcal{X}$.
Hence, the GFP problem is
\begin{equation}\label{22}
\begin{aligned}
 & \max_{\mathbf{x}}\min_{p=1,...,P}\frac{f_{p}\left(\mathbf{x}\right)}{g_{p}\left(\mathbf{x}\right)} \\
 & \mathrm{s.t.}~ \mathbf{x}\in\mathcal{X}.
\end{aligned}
\end{equation}
By introducing an auxiliary variable $t$, the equivalent quadratic transformation of the max-min-ratio problem~\eqref{22} is expressed as
\begin{equation}\label{23}
\begin{aligned}
 & \max_{\mathbf{x},\mathbf{y},t}~t  \\
 & \mathrm{s.t.}~ 2y_{p}\sqrt{f_{p}\left(\mathbf{x}\right)}-y_{p}^{2}g_{p}\left(\mathbf{x}\right) \geq t, \mathbf{x}\in\mathcal{X},y_{p}\in\mathbb{R},
\end{aligned}
\end{equation}
where $\mathbf{y}$ is a collection of variables $\{y_1,...,y_P\}$.
When $\mathbf{x}$ is held fixed, the optimal $y_{p}$ can be found in closed form as $y_p^\star=\frac{\sqrt{f_p\left(\mathbf{x}\right)}}{g_p\left(\mathbf{x}\right)}, \forall p$.
\end{lemma}
\textbf{Remark:} The GFP Algorithm is characterized by a linear convergence rate and, in each iteration, only requires the solution of a convex problem, which can be obtained in polynomial time using many convex programming algorithms.
Furthermore, the objective function of~\eqref{22} monotonically converges to the optimal value of~\eqref{22}.}

Based on the sequential optimization framework,~\eqref{P1} belongs to the class of GFP problems.
Hence, by introducing an auxiliary variable $t$,~\eqref{P1} can be reformulated as
\begin{subequations}\label{P2}
 \begin{align}
 &\underset{\mathbf{x},\mathbf{\Theta},{{\eta }},t }{\mathop{\max }}\,  t +\sum\nolimits_{p=1}^{P}\omega \left( \sum\nolimits_{n=1}^{N}{{{\eta }_{n,p}^2}}-\sum\nolimits_{n=1}^{N}{\eta _{n,p}} \right) \label{P21}  \\
&\text{s.t.}~~~ g_p \ge t, \forall p, \label{P26}\\
&~~~~~~~\eqref{P12},~\eqref{P15}, \label{P25}
\end{align}
\end{subequations}
where
\begin{equation}\label{aaa1}
\begin{aligned}
g_p=2{u}_{p}\operatorname{Re}\left( \mathbf{w}^{H}_p{\mathbf{A}}_{p}\mathbf{x} \right)
-u_{p}^{2}\mathbf{w}^{H}_p\left( {{\mathbf{\Pi }}_{c}}+\mathbf{\Sigma }_p+\mathbf{I}\sigma _{R}^{2} \right)\mathbf{w}_p. \\
\end{aligned}
\end{equation}
To solve the reformulated problem~\eqref{P2}, $\eta _{n,p}$ is updated until convergence and then the optimal solutions $\mathbf{x, \Theta}$ can be found by updating $u_{p},\forall p$ iteratively by
\begin{equation}\label{uim1}
u_{p}^{(l+1)}= \frac{\mathbf{w}^{H}_p\mathbf{A}_{p}^{(l)} \mathbf{x}^{(l)}}
{\mathbf{w}^{H}_p\left( {\mathbf{\Pi }}_{c}^{(l)}+\mathbf{\Sigma }_p^{(l)}+\mathbf{I}\sigma _{R}^{2} \right)\mathbf{w}_p},\forall p.
\end{equation}

\subsection{Solve for ${\eta}$ and $\mathbf{x}$}

Let $g\left( \eta _{n,p} \right)=\sum\nolimits_{n=1}^{N}{\eta _{n,p}^{2}}$, we give the first-order Taylor expansion of $g\left( \eta _{n,p} \right)$ as follows:
\begin{equation}\label{bargeta}
\color{blue}\bar{g}\left( \eta _{n,p},\eta _{n,p}^{\left( l \right)} \right)\ge \sum\limits_{n=1}^{N}{{{\left( \eta _{n,p}^{\left( l \right)} \right)}^{2}}}+2\sum\limits_{n=1}^{N}{\eta _{n,p}^{\left( l \right)}}\left( \eta _{n,p}-\eta _{n,p}^{\left( l \right)} \right).
\end{equation}
Fixed other variables, problem~\eqref{P2} is reduced to
\begin{subequations}\label{eta}
\begin{align}
  & \underset{\eta }{\mathop{\min }}\, \sum\nolimits_{p=1}^{P}\omega \left( \sum\nolimits_{n=1}^{N}{\eta _{n,p}}-\bar{g}  \left(\eta _{n,p},\eta _{n,p}^{\left( l-1 \right)} \right) \right) \label{eta1}  \\
 &\text{s}\text{.t}\text{.   }~\eqref{barhtn31},~\eqref{barhtn32},~0 \le \eta _{n,p} \le 1,\forall n,p. \label{eta2}
\end{align}
\end{subequations}
Clearly, the optimal $\eta_{n,p}$ can be obtained by solving~\eqref{eta} until $\left| \sum\nolimits_{n=1}^{N}{\eta _{n,p}^{\left( l-1 \right)}}\left( \eta _{n,p}-\eta _{n,p}^{\left( l-1 \right)} \right) \right|<{{\varepsilon }_{\eta }}$, with ${{\varepsilon }_{\eta }}$ being a given convergence threshold.
%By defining $\mathbf{z }_0={{\left[ {\mathbf{z }_0^{T}}\left( 1 \right),\ldots,{\mathbf{z }_0^{T}}\left( N \right) \right]}^{T}}\in {{\mathbb{C}}^{{{N}_{I}}N\times 1}}$,
%The receive signal at the hybrid RIS is given by ${{\mathbf{r}}_{R}}=\left( \mathbf{e}_{n}^{T}\otimes \mathbf{G} \right)\mathbf{x}+\mathbf{z}_0(n)$.
%Define $\mathbf{U}={{\left( \mathbf{e}_{n}^{T}\otimes \mathbf{G} \right)}^{H}}{{\mathbf{\Psi }}^{H}}\mathbf{\Psi }\left( \mathbf{e}_{n}^{T}\otimes \mathbf{G} \right)$ and $\bar{P}_{\text{max}}^{\text{R}}=P_{\text{max}}^{\text{R}}-\sigma _{\text{ris}}^{2}\text{tr}\left[ \mathbf{\Psi }{{\mathbf{\Psi }}^{H}} \right]$ then
Hence, fixed other variables, the waveform optimization subproblem can be cast as
\begin{subequations}\label{px}
 \begin{align}
& \underset{\mathbf{x}}{\mathop{\max }}\,~t \label{px1} \\
&~\text{s.t. }~~g_p\left( \mathbf{x} \right)\ge t, \forall p, \label{P26a}\\
& ~~~~~~\eqref{P02},~\eqref{P03},~\eqref{P15}. \label{px4}
\end{align}
\end{subequations}
where
$g_p\left( \mathbf{x} \right)=2{u}_{p}\operatorname{Re}\left( \mathbf{w}^{H}_p{{\mathbf{A}}_{p}}\mathbf{x} \right)-u_{p}^{2}\left(\mathbf{x}^{H}\mathbf{C}_p\mathbf{x} +c_p\right)$,
${{\mathbf{C}}_{p}}=\bar{\mathbf{P}}^H(\nu_1)\mathbf{\bar{C}}^H{{\mathbf{\bar{H}}}_{\text{c}}^H}\mathbf{w}_p\mathbf{w}_p^H{{\mathbf{\bar{H}}}_{\text{c}}}\mathbf{\bar{C}}\bar{\mathbf{P}}(\nu_1)$,
$c_p=\mathbf{w}_p^{H}\left( \mathbf{\Sigma }_p+\mathbf{I}\sigma _{R}^{2} \right)\mathbf{w}_p$.
Clearly, the constraint~\eqref{P26a} is a concave function w.r.t. $\mathbf{x}$.
Hence, problem~\eqref{px} is convex and can be solved by convex optimization solvers, such as CVX.

\subsection{Solve for $\mathbf{\Theta }$}

{\color{blue}This subsection discusses the multiple hybrid RIS beamforming design with DRC.
The optimization problem becomes a combinatorial optimization problem.
Though the optimized solution can be obtained via an exhaustive search, it requires a prohibitive complexity since the number of units on the multiple RIS is usually large.
To overcome this drawback, we make a small modification on that two-step method:
Firstly, we handle the optimization problem by relaxing the DRC constraint~\eqref{P04} to a continuous reflection coefficient (CRC) constraint, i.e.,
\begin{equation}\label{P041}
\left| {{\vartheta }_{s,n}} \right|=1, n\notin \mathbb{Q},
\end{equation}
and derive the optimal $\mathbf{\Theta}$.
Secondly, $\mathbf{\Theta }$ is projected to the nearest feasible point in the discrete constraint set~\eqref{P04} to find a suboptimal solution for the non-convex DRC optmization problem.}

Given the other variables, the problem of tackling the weights of the hybrid RIS is difficult owing to the presence of the quartic term in the sensing SINR function.
This high-order term actually derives from the transmission channel of the sensing signals, which produces the quartic term in the objective function.
To overcome this challenge, we decompose the high-order term, for example, reducing the sensing SINR function from quartic to quadratic, which makes it easier to deal with.
{\color{blue}Let $\boldsymbol{\vartheta}_s\triangleq{{\left[ {\vartheta _{s,1}},\cdots ,{\vartheta _{s,{N}_{I}}} \right]}^{T}}$ be the vector formed by the diagonal elements of $\mathbf{\Theta}_s$ and define the vector $\boldsymbol{\vartheta}=[1,\boldsymbol{\vartheta}_1^T,\cdots,\boldsymbol{\vartheta}_S^T]^T$.
Then, by introducing a copy $\boldsymbol{\vartheta }_1$ of $\boldsymbol{\vartheta }$, problem~\eqref{P2} in an explicit expression w.r.t. $\boldsymbol{\vartheta }$ and $\boldsymbol{\vartheta }_1$ with fixed other variables are formulated as
\begin{subequations}\label{Pvartheta}
\begin{align}
&\underset{\boldsymbol{\vartheta}, {\boldsymbol{\vartheta }_{1}} }{\mathop{\max }}\, ~t \\
&\text{s.t.}~ h_p\left( \boldsymbol{\vartheta},{\boldsymbol{\vartheta }_{1}}  \right)\geq t, \forall p, \\
&~~~~{\boldsymbol{\vartheta }_{1}}=\boldsymbol{\vartheta }, \\
&~~~~\eqref{P04},~\eqref{PR1},~\eqref{CI11},~\eqref{barht3},~\eqref{barht4}, \label{Pvartheta10}
\end{align}
\end{subequations}
where $h\left( \boldsymbol{\vartheta},{\boldsymbol{\vartheta }_{1}}  \right)$ is given by~\eqref{hvartheta11} at the top of the next page.
The equivalence between problem~\eqref{P2} and problem~\eqref{Pvartheta} is proved in Appendix~\ref{3f}, in which the expressions of ${\mathbf{D}}_{\text{e},p}$, ${\mathbf{D}}_{\text{c},p}$  ${\mathbf{F}}_{p}$, ${\mathbf{\hat H}}_{k}$, $\bar{\mathbf{\Lambda}}_k$, ${\mathbf{\hat H}}_{e}^{p}$ and $\bar{\mathbf{B}}_e^p$ are shown in~\eqref{DDF},~\eqref{hatHk},~\eqref{Lambdak} and~\eqref{Bep}.
\begin{figure*}
\begin{equation}\label{hvartheta11}
h_p\left( \boldsymbol{\vartheta} ,{\boldsymbol{\vartheta }_{1}} \right)=
2{{u}_{p}}\operatorname{Re} \left( \mathbf{w}_{p}^{H}{{\mathbf{D}}_{\text{e},p}}\text{vec}\left\{ {\boldsymbol{\vartheta}}{\boldsymbol{\vartheta }_{1}^{T}} \right\}  \right)
-u_{p}^{2}\text{vec}\left\{ {\boldsymbol{\vartheta }^{*}}{\boldsymbol{\vartheta }_{1}^{T}} \right\} ^H{{\mathbf{D}}_{\text{c},p}}\text{vec}\left\{ {\boldsymbol{\vartheta }^{*}}{\boldsymbol{\vartheta }_{1}^{T}} \right\}-u_{p}^{2}\boldsymbol{\vartheta }^{H}{{\mathbf{F}}_{p}}\boldsymbol{\vartheta }-u_{p}^{2}\sigma _{R}^{2}\mathbf{w}_p^{H}\mathbf{w}_p,
\end{equation}
\end{figure*}
\begin{figure*}
\begin{equation}\label{PR1}
\begin{aligned}
&{{P}_{R,s}}=\boldsymbol{\vartheta}^H_s\mathcal{I}_{{{N}_{I}}}^{\mathbb{Q}}\text{diag}\left( {{\mathbf{x}}^{H}}{{\left( \mathbf{e}_{n}^{T}\otimes \mathbf{G}_s \right)}^{H}}\right) \text{diag}\left( {{\mathbf{x}}^{H}}{{\left( \mathbf{e}_{n}^{T}\otimes \mathbf{G}_s \right)}^{H}}\right)\mathcal{I}_{{{N}_{I}}}^{\mathbb{Q}}\boldsymbol{\vartheta}_s
+\sigma _{\text{RIS}}^{2}\boldsymbol{\vartheta}^H_s\mathcal{I}_{{{N}_{I}}}^{\mathbb{Q}}\mathcal{I}_{{{N}_{I}}}^{\mathbb{Q}}\boldsymbol{\vartheta}_s, \forall s.
\end{aligned}
\end{equation}
\end{figure*}
\begin{figure*}
\begin{equation}\label{CI11}
\begin{aligned}
&\left| \operatorname{Im}(\boldsymbol{\vartheta}^T\hat{\mathbf{H}}_{k}\mathbf{x}\left(n \right)s_{k}^{*}  \right|\le \left(\operatorname{Re}(\boldsymbol{\vartheta}^T\hat{\mathbf{H}}_{k}\mathbf{x}\left(n \right)s_{k}^{*} )
-\sqrt{{\Gamma }_{k}\left(\sigma _\text{RIS}^{2}\boldsymbol{\vartheta }^H\bar{\mathbf{\Lambda}}_k\boldsymbol{\vartheta }+\sigma _{k}^2\right)} \right)\tan \phi_k ,\forall k, n,
\end{aligned}
\end{equation}
\end{figure*}
\begin{figure*}
\begin{subequations}\label{DI11}
\begin{align}
&\color{blue}\left( \operatorname{Re}\left(\boldsymbol{\vartheta}^T {\mathbf{\hat H}}_{\mathrm{e}}^{p}\mathbf{x}\left(n\right) s_{k}^{*} \right)
-\sqrt{{\Gamma }_{\text{E}}\left(\sigma _\text{RIS}^{2}\boldsymbol{\vartheta }^H\bar{\mathbf{B}}_e^p \boldsymbol{\vartheta}+\sigma _{e}^2\right)}\right)\tan {{\phi }_{\text{e}}}
-\operatorname{Im}\left( \boldsymbol{\vartheta}^T {\mathbf{\hat H}}_{\mathrm{e}}^{p}\mathbf{x}\left(n\right) s_{k}^{*}  \right)-{{\eta }_{n,p}}\Upsilon \le 0,\forall n,p,  \label{barht3} \\
&\color{blue}\left( \operatorname{Re}\left(\boldsymbol{\vartheta}^T {\mathbf{\hat H}}_{\mathrm{e}}^{p}\mathbf{x}\left(n\right) s_{k}^{*} \right)
-\sqrt{{\Gamma }_{\text{E}}\left(\sigma _\text{RIS}^{2}\boldsymbol{\vartheta }^H\bar{\mathbf{B}}_e^p \boldsymbol{\vartheta}+\sigma _{e}^2\right)}\right)\tan {{\phi }_{\text{e}}}
+\operatorname{Im}\left(\boldsymbol{\vartheta}^T{\mathbf{ \hat H}}_{\mathrm{e}}^{p}\mathbf{x}\left(n\right) s_{k}^{*}  \right)-(1-{{\eta }_{n,p}})\Upsilon \le 0,\forall n,p, \label{barht4}
\end{align}
\end{subequations}
\hrulefill
\end{figure*}

1) \textbf{Optimization Step:}
By exploiting the PDD technique~\cite{Shi2020} and relaxing the DRC constraint~\eqref{P04} to the CRC constraint~\eqref{P041}, the optimized solution of problem~\eqref{Pvartheta} can be efficiently found by iteratively optimizing its augmented Lagrangian problem, i.e.,
\begin{subequations}\label{Pvartheta1}
\begin{align}
& \underset{\boldsymbol{\vartheta}, {\boldsymbol{\vartheta }_{1}} }{\mathop{\min }}\,-t
+\frac{\rho }{2}\left\| \boldsymbol{\vartheta} -{\boldsymbol{\vartheta }_{1}} \right\|_{2}^{2}+\operatorname{Re} \left\{ {{\mathbf{\lambda }}^{H}}\left( \boldsymbol{\vartheta } -{\boldsymbol{\vartheta }_{1}} \right) \right\} \\
&\text{s.t. }~~h_p\left( \boldsymbol{\vartheta},{\boldsymbol{\vartheta }_{1}}  \right)\geq t, \forall p,  \label{hpvartheta}   \\
& ~~~~~~\eqref{P041},~\eqref{PR1},~\eqref{CI11},~\eqref{barht3},~\eqref{barht4}. \label{Pvartheta101}
\end{align}
\end{subequations}}
Then, we design a two-layer iteration process based on the PDD framework.
In particular, $\boldsymbol{\vartheta}$ and $\boldsymbol{\vartheta}_1$ are updated employing block coordinate descent (BCD) method within its inner layer.
Then, the penalty factor $\rho$ or the dual variable $\mathbf{\lambda}$ is selectively updated within its outer layer.
For the inner layer iteration, we update $\boldsymbol{\vartheta}$ and $\boldsymbol{\vartheta}_1$ iteratively.
Fixing $\boldsymbol{\vartheta}_1$, the maximization of augmented Lagrangian w.r.t. $\boldsymbol{\vartheta}$ reduces to solving the following problem
\begin{subequations}\label{Pvartheta2}
 \begin{align}
 &\underset{\boldsymbol{\vartheta} }{\mathop{\min }}\, -t
+\frac{\rho }{2}\left\| \boldsymbol{\vartheta} -{\boldsymbol{\vartheta }_{1}} \right\|_{2}^{2}+\operatorname{Re} \left\{ {{\mathbf{\lambda }}^{H}}\left( \boldsymbol{\vartheta } -{\boldsymbol{\vartheta }_{1}} \right) \right\} \\
& \text{s.t.}~~~h_p\left( \boldsymbol{\vartheta}  \right)\geq t, \forall p, \label{1a}\\
& ~~~~~~\eqref{Pvartheta101}.
\end{align}
\end{subequations}
{\color{blue}It can be observed that the DI-type constraints~\eqref{DI11} are non-convex due to the presence of the term $\sqrt{{\Gamma }_{\text{E}}\left(\sigma _\text{RIS}^{2}\boldsymbol{\vartheta }^H\bar{\mathbf{B}}_e^p \boldsymbol{\vartheta}+\sigma _{e}^2\right)}$.
By applying the first-order Taylor series expansion of $\sqrt{{\Gamma }_{\text{E}}\left(\sigma _\text{RIS}^{2}\boldsymbol{\vartheta }^H\bar{\mathbf{B}}_e^p \boldsymbol{\vartheta}+\sigma _{e}^2\right)}$ at the point $\boldsymbol{\vartheta }^{(l)}$,~\eqref{barht3} and~\eqref{barht4} can be recast as
\begin{subequations}\label{DI1t}
\begin{align}
&\color{blue}\left( \operatorname{Re}\left(\boldsymbol{\vartheta}^T{\mathbf{\hat H}}_{\mathrm{e}}^{p}\mathbf{x}\left(n\right) s_{k}^{*} \right)-f^p\left(\boldsymbol{\vartheta }\right)\right)\tan {{\phi }_{\text{e}}} \nonumber \\
&-\operatorname{Im}\left( \boldsymbol{\vartheta}^T{\mathbf{ \hat H}}_{\mathrm{e}}^{p}\mathbf{x}\left(n\right) s_{k}^{*} \right)
-{{\eta }_{n,p}}\Upsilon \le 0,\forall n,p,  \label{DI1t1} \\
&\color{blue}\left( \operatorname{Re}\left(\boldsymbol{\vartheta}^T {\mathbf{\hat H}}_{\mathrm{e}}^{p}\mathbf{x}\left(n\right) s_{k}^{*} \right)-f^p\left(\boldsymbol{\vartheta }\right)\right)\tan {{\phi }_{\text{e}}} \nonumber  \\
&+\operatorname{Im}\left(\boldsymbol{\vartheta}^T \hat{\mathbf{H}}_{\mathrm{e}}^{p}\mathbf{x}\left(n\right) s_{k}^{*} \right)-(1-{{\eta }_{n,p}})\Upsilon \le 0,\forall n,p, \label{DI1t2}
\end{align}
\end{subequations}
where
\begin{equation}
\begin{aligned}
&f_p\left(\boldsymbol{\vartheta }\right)
=\sqrt{\Gamma_\mathrm{E}\left(\sigma_\mathrm{RIS}^2\boldsymbol{\vartheta}^{(l)H}\bar{\mathbf{B}}_e^p\boldsymbol{\vartheta}^{(l)}+\sigma_e^2\right)}  \\
&+\frac{\Gamma_\mathrm{E}\sigma_\mathrm{RIS}^2\text{Re}\left\{\boldsymbol{\vartheta}^{(l)H}\bar{\mathbf{B}}_e^p(\boldsymbol{\vartheta}-\boldsymbol{\vartheta}^{(l)})\right\}}
{\sqrt{\Gamma_\mathrm{E}\left(\sigma_\mathrm{RIS}^2\boldsymbol{\vartheta}^{(l)H}\bar{\mathbf{B}}_e^p\boldsymbol{\vartheta}^{(l)}+\sigma_e^2\right)}}, \forall p.
\end{aligned}
\end{equation}
It is evident that $f_p\left(\boldsymbol{\vartheta }\right)$ is a concave function w.r.t. $\boldsymbol{\vartheta }$.}
The unit-modulus constraint of~\eqref{P041} is also non convex.
Based on the PCCP framework~\cite{2016thomas}, the unit-modulus constraint of passive RIS units can be rewritten as $1\le {{\left| {{\vartheta }_{s,n}} \right|}^{2}}\le 1,n\notin \mathbb{Q}, \forall s$.
The non-convex part of the equivalent constraint is then linearized by $1\le 2\operatorname{Re}\left( \vartheta_{s,n}^{(l-1)}\bar{\vartheta }_{s,n}^{*} \right)-{{\left| \vartheta_{s,n}^{(l-1)} \right|}^{2}},n\notin \mathbb{Q}, \forall s,$ at fixed $\vartheta_{s,n}^{(l-1)}$.
Thus, introducing slack variables $\mathbf{e}={{\left[ {{e}_{1}},\ldots ,{{e}_{2{{N}_{I}}}} \right]}^{T}}$, we have the convex subproblem of solving $\boldsymbol{\vartheta}$ as
\begin{subequations}\label{PBphi}
\begin{align}
& \color{blue}\underset{\boldsymbol{\vartheta} ,\mathbf{e}}{\mathop{\min }}\,-t
+\frac{\rho }{2}\left\| \boldsymbol{\vartheta} -{\boldsymbol{\vartheta }_{1}} \right\|_{2}^{2}+\operatorname{Re} \left\{ {{\mathbf{\lambda }}^{H}}\left( \boldsymbol{\vartheta } -{\boldsymbol{\vartheta }_{1}} \right) \right\} \nonumber \\
&~~~~~~ -{{\kappa }^{\left( l \right)}}\sum\nolimits_{n=1}^{2{N_I}}{{e}_{n}} \\
&\text{s.t.}~h_p\left( \boldsymbol{\vartheta}  \right)\geq t, \forall p,  \\
&~~~~~{{\left| \vartheta _{s,n}^{(l-1)} \right|}^{2}}-2\operatorname{Re} \left( \vartheta _{s,n}^{(l-1)}\vartheta_{s,n}^{*} \right)\le e_{n}-1, \\
&~~~~~{{\left| {{ \vartheta}_{s,n}} \right|}^{2}}\le 1+e_{{n+N_I}}, \\
&~~~~~\eqref{PR1},~\eqref{CI11},~\eqref{DI1t1},~\eqref{DI1t2},~\mathbf{e}\ge 0,
\end{align}
\end{subequations}
where the penalty term ${{\left\| \mathbf{e} \right\|}_{1}}$ is scaled by the regularization factor ${{\kappa }^{\left( l \right)}}$ to control the feasibility of the constraints.
\eqref{PBphi} is a second-order cone programming (SOCP) problem and thus can be solved easily by CVX.
The PCCP-based scheme to solve~\eqref{PBphi} is summarized in Algorithm~\ref{alg1}.
We remark that the unit-modulus constraint in~\eqref{Pvartheta1} is ensured by ${{\left\| \mathbf{e} \right\|}_{1}}\le \chi $ when $\chi$ is sufficiently small.
The coefficient ${\kappa }_\text{max}$ is enforced to evade a numerical problem, i.e., a solution meeting ${{\left\| \mathbf{e} \right\|}_{1}}\le \chi $ may not be obtained if the iteration procedure converges to the exit condition $\left\| {\boldsymbol{\vartheta }^{(l)}}-{\boldsymbol{\vartheta}^{(l-1)}} \right\|_{1}>{{\zeta }_{1}}$ with the increase of ${{\kappa }^{(t)}}$.
\begin{algorithm}[t]
	\caption{PCCP procedure for solving~\eqref{PBphi}}
    \label{alg1}
 \hspace*{0.02in} {\bf  Initialize:} $\boldsymbol{\vartheta}^{\left( 0 \right)}$, ${{\gamma }^{\left( 0 \right)}}>1$ and $l=0$; \\
 \vspace{-0.4cm}
	\begin{algorithmic}[1]		
\State	\textbf{repeat} \\
\ \  \textbf{if} $l<L_\text{max}$ \textbf{then} \\
\ \  obtain $\boldsymbol{\vartheta}^{\left( l+1 \right)}$ by solving~(\ref{PBphi}); \\  %$\boldsymbol{\phi}^{\left( t+1 \right)}$
\ \  ${{\kappa }^{(l+1)}}=\min \left\{ \gamma {{\kappa }^{(l)}},{{\kappa }_{\max }} \right\}$; \\
\ \  $l=l+1$; \\
\ \  \textbf{else} \\
\ \   Initialize $\boldsymbol{\vartheta}^{\left( 0 \right)}$, ${{\gamma }^{\left( 0 \right)}}>1$ and  $l=0$; \\
\ \  \textbf{end if} \\
\textbf{until} ${{\left\| \mathbf{e} \right\|}_{1}}\le \chi $ and $\left\| {\boldsymbol{\vartheta }^{(l)}}-{\boldsymbol{\vartheta }^{(l-1)}} \right\|_{1}>{{\zeta }_{1}}$. \\
\textbf{Output} ${\boldsymbol{\vartheta }^{(m)}}={\boldsymbol{\vartheta }^{(l)}}$.
	\end{algorithmic}
\end{algorithm}
Given the variable $\boldsymbol{\vartheta}$, the problem of optimizing the auxiliary copy variable $\boldsymbol{\vartheta}_1$ is recast as
\begin{subequations}
\begin{align}\label{Pvartheta3}
&\color{blue}\underset{\boldsymbol{\vartheta}_1 }{\mathop{\min }}\,-t+\frac{\rho }{2}\left\| \boldsymbol{\vartheta} -{\boldsymbol{\vartheta }_{1}} \right\|_{2}^{2}+\operatorname{Re} \left\{ {{\mathbf{\lambda }}^{H}}\left( \boldsymbol{\vartheta }-{\boldsymbol{\vartheta }_{1}} \right) \right\},\\
&\color{blue}\text{s.t.}~~h_p\left( \boldsymbol{\vartheta}_1 \right)\geq t, \forall p,
\end{align}
\end{subequations}
The problem of solving $\boldsymbol{\vartheta}_1$ is convex and hence can be efficiently solved by the CVX directly.
Since ${\boldsymbol{\vartheta }}$ and ${\boldsymbol{\vartheta }_{1}}$ are updated using BCD method within the inner layer iteration, the objective value of~\eqref{Pvartheta1} converges to the limit value.
When its convergence is achieved, we regulate $\mathbf{\lambda }$ or $\rho$ within the outer layer iteration.
Precisely, if ${{\left\| \boldsymbol{\vartheta } -{\boldsymbol{\vartheta }_{1}} \right\|}_{\infty }}$ is smaller than the predefined threshold~\cite{Shi2020}, the coefficient $\mathbf{\lambda }$ is updated in a gradient ascent way, i.e., ${{\mathbf{\lambda }}^{(m+1)}}={{\mathbf{\lambda }}^{(m)}}+\rho \left( \boldsymbol{\vartheta }-{\boldsymbol{\vartheta }_{1}} \right)$.
If the equality constraint $\boldsymbol{\vartheta } ={\boldsymbol{\vartheta }_{1}}$ is not guaranteed, to impose $\boldsymbol{\vartheta } ={\boldsymbol{\vartheta }_{1}}$ being obtained in the subsequent iterations, the control factor $\rho$ is updated in the outer layer procedure, i.e., ${{\rho }^{\left( m+1 \right)}}={{c}_{0}}{{\rho }^{\left( m \right)}}$, with ${{c}_{0}}$ being a positive constant.
The proposed PDD procedure for solving~\eqref{Pvartheta1} is presented in Algorithm~\ref{alg2}.

{\color{blue}2) \textbf{Projection Step:}
Denote the optimized solution $\boldsymbol{\vartheta}$ for the case when the elements in $\boldsymbol{\vartheta}$ follows the unit modulus constraint.
Then, we can directly quantize the optimized solution $\boldsymbol{\vartheta}$ in the CRC case to the nearest feasible point
$\boldsymbol{\hat \vartheta}^{*}$ in the DRC case.
Specifically, the preliminary solution $\boldsymbol{\hat \vartheta}^{*}$ for~\eqref{Pvartheta} can be generated by the following formula:\par
\begin{footnotesize}
\begin{equation}\label{1a}
\begin{aligned}
  & \frac{{{\psi }_{n}}}{\left| {{a}_{n}} \right|}=\left\{ \begin{matrix}
   1, & \text{if }\arg \left( \frac{{{\left[ \boldsymbol{\vartheta}  \right]}_{n}}}{\left| {{a}_{n}} \right|} \right)\in \left[ \frac{-\pi }{{{2}^{b}}},\frac{\pi }{{{2}^{b}}} \right),  \\
   \vdots  & \vdots   \\
   {{e}^{j\frac{m2\pi }{{{2}^{b}}}}}, & \text{if }\arg \left( \frac{{{\left[ \boldsymbol{\vartheta}\right]}_{n}}}{\left| {{a}_{n}} \right|} \right)\in \left[ \frac{(2m-1)\pi }{{{2}^{b}}},\frac{(2m+1)\pi }{{{2}^{b}}} \right),  \\
   \vdots  & \vdots   \\
   {{e}^{j\frac{({{2}^{b}}-1)2\pi }{{{2}^{b}}}}}, \!&\! \text{if}\arg \!\left(\! \frac{{{\left[ \boldsymbol{\vartheta}  \right]}_{n}}}{\left| {{a}_{n}} \right|}\! \right)\!\in \!\left[\! \frac{({{2}^{b+1}}-3)\pi }{{{2}^{b}}},\frac{({{2}^{b+1}}-1)\pi }{{{2}^{b}}}\! \right)\!,  \\
\end{matrix} \right. \\
 & n\in \mathbb{Q}, \\
\end{aligned}
\end{equation}
\end{footnotesize}%
\begin{footnotesize}
\begin{equation}\label{2a}
\begin{aligned}
 & {{\phi }_{n}}=\left\{ \begin{matrix}
1, & \text{if } \arg\left(\left[\boldsymbol{\vartheta}\right]_n\right) \in \left[\frac{-\pi}{2^b}, \frac{\pi}{2^b}\right), \\
\vdots & \vdots \\
e^{j\frac{m2\pi}{2^b}}, & \text{if } \arg\left(\left[\boldsymbol{\vartheta}\right]_n\right) \in \left[\frac{(2m-1)\pi}{2^b}, \frac{(2m+1)\pi}{2^b}\right), \\
\vdots & \vdots \\
e^{j\frac{(2^b-1)2\pi}{2^b}}, \!&\! \text{if } \arg\left(\left[\boldsymbol{\vartheta}\right]_n\!\right) \!\in\! \left[\frac{(2^{b+1}-3)\pi}{2^b}, \frac{(2^{b+1}-1)\pi}{2^b}\right),  \\
\end{matrix} \right. \\
& n\notin \mathbb{Q},\\
\end{aligned}
\end{equation}
\end{footnotesize}%
However, the preliminary solution $\boldsymbol{\hat  \vartheta}^{*}$ obtained by projection scheme may be not a locally optimal solution.
In order to make the objective function's value to be non-decreasing after each iteration for $\boldsymbol{\hat \vartheta}^{*}$, we update $\boldsymbol{\hat \vartheta}^{*}$ only when $\underset{\theta^p,{\varphi}^p}{\mathop{\min }}\text{SINR}_{p}\left(\boldsymbol{\hat \vartheta}^{*(l)}\right) \geq \underset{\theta^p,{\varphi}^p}{\mathop{\min }}\text{SINR}_{p}\left(\boldsymbol{\hat \vartheta}^{*(l-1)}\right)$.
The GFP-based iterative algorithm for optimizing the hybrid RIS-aided ISAC secure system is specified in Algorithm~\ref{alg3}.}

\begin{algorithm}[t]
	\caption{PDD procedure for solving~\eqref{Pvartheta1}}
    \label{alg2}
    \hspace*{0.02in} {\bf  Initialize:} ${\boldsymbol{\vartheta }^{(0)}}, \boldsymbol{\vartheta} _{1}^{(0)}, {{\mathbf{\lambda }}^{(0)}}, {{\rho }^{(0)}}$, and $m=0$; \\
    \vspace{-0.4cm}
	\begin{algorithmic}[1]		
\State	\textbf{repeat} \\
\ \  set ${\boldsymbol{\vartheta }^{(m,0)}}={\boldsymbol{\vartheta }^{(m)}}$, $\boldsymbol{\vartheta} _{1}^{(m,0)}=\boldsymbol{\vartheta} _{1}^{(m)}$, and $i=0$; \\
\ \   \textbf{repeat} \\
\ \  update ${\boldsymbol{\vartheta }^{(m,i+1)}}$ using Algorithm~\ref{alg1};  \\
\ \  $i=i+1$; \\
\ \  \textbf{until} convergence \\
\ \  set ${\boldsymbol{\vartheta }^{(m+1)}}={\boldsymbol{\vartheta }^{(m,\infty )}},\boldsymbol{\vartheta} _{1}^{(m+1)}=\boldsymbol{\vartheta} _{1}^{(m,\infty )}$; \\
\ \  \textbf{if} ${{\left\| {\boldsymbol{\vartheta }^{(m+1)}}-\boldsymbol{\vartheta}_{1}^{(m\text{+}1)} \right\|}_{\infty }}\le {{\varpi }_{1}}$ \textbf{then} \\
\ \   update $\mathbf{\lambda }$ and ${{\rho }^{(m+1)}}={{\rho }^{(m)}}$; \\
\ \   update ${{\rho }^{(m+1)}}$ and ${{\mathbf{\lambda }}^{(m+1)}}={{\mathbf{\lambda }}^{(m)}}$; \\
\ \  \textbf{end if} \\
\ \  $m=m+1$; \\
\textbf{until} ${{\left\| {\boldsymbol{\vartheta }^{(m+1)}}-\boldsymbol{\vartheta}_{1}^{(m\text{+}1)} \right\|}_{\infty }}\le {{\varpi }_{1}}$.
	\end{algorithmic}
\end{algorithm}
The computational complexity of Algorithm~\ref{alg3} is determined by the number of outer iterations $\bar{N}$, and the complexity required at each iteration.
Exactly, it involves the complexity of ${{\left( {{\mathbf{\Pi }}_{c}}+\mathbf{\Sigma }_p+\sigma _{R}^{2}\mathbf{I} \right)}^{-1}}$, the complexity of solving~\eqref{px} and the complexity of Algorithm~\ref{alg2}.
The complexities for updating the receive filter bank $\left\{ \mathbf{w}_{p} \right\}_{p=1}^{P}$ and transmit waveform $\mathbf{x}$ are given as $\mathcal{O}\left( P{{\left( M{{N}_{a}} \right)}^{3}} \right)$ and $\mathcal{O}\left( P\left( MNN_{a}^{2}+{{M}^{2}}N_{a}^{2} \right) \right)$ by reserving the highest order term.
The complexity for updating the SCUs' receive beamformers $\left\{ \mathbf{v}_{k} \right\}_{k=1}^{K}$ is given as $\mathcal{O}\left( KN^3_r \right)$.
For Algorithm~\ref{alg2}, the classical interior point approach is used to solve~\eqref{PBphi}, whose complexity is relevant to the dimension of the variable and the number of linear matrix inequality (LMI) constraints and second-order cone (SOC) constraints.
Thus, the complexity of Algorithm~\ref{alg2} is approximately expressed by
\begin{footnotesize}$\mathcal{O}\left( {{T}_\text{in}}{{T}_\text{out}}\left( P{S{N}_{I}}{{M}^{2}}N_{a}^{2}+P{{\left( \left( 3N+KN \right)S{{N}_{I}} \right)}^{{1}/{2}}}2KN(SN^4_{I}) \right) \right)$\end{footnotesize}, where ${{T}_\text{in}}$ and ${{T}_\text{out}}$ are the number of the outer and inner PDD iterations, respectively.

\section{Numerical Results}\label{simulation}

In this section, simulation results are offered to demonstrate the validity of the devised design for hybrid RIS-enhanced ISAC secure systems.
Unless otherwise specifically mentioned, the following simulation parameters are employed.
We consider that the BS equipped with $N_a=8$ transmit/receive antennas sends dual-functional waveforms to serve $K=3$ SCUs and sense an Eve.
{\color{blue}Each SCU is equipped with $N_r=4$ receive antennas.}
The TIR support interval is considered as $Q=17$.
%The sampling interval of the TAA is ${{0.1}^{\circ }}$.
We calculate the actual TIR using the toolbox presented in~\cite{2015MaioDe}.
The maximum power budget of the BS and hybrid RIS are $P_{\max }^{\text{B}}=30\,\,\text{dBm}$ and $P_{\max }^{\text{R}}=20\,\,\text{dBm}$.
The total number of hybrid RIS units is $N_I=30$ and the number of active reflection units on the hybrid RIS is $A=6$.
The SINR thresholds for the SCUs and the Eve are set as ${{\Gamma }_{k}}=10\,\,\text{dB}, \forall k$ and ${{\Gamma }_{E}}=-1\,\,\text{dB}$, respectively.
We set the additive noise power at the BS and each SCU as $\sigma _{R}^{2}=\sigma _{C,k}^{2}=-80\,\,\text{dBm}, \forall k$.
The active reflecting unit noise power is $\sigma _\text{ris}^{2}=-70\,\,\text{dBm}$.
%${\theta }_{p},{\varphi}_{p}, \forall p$
\begin{algorithm}[t]
	\caption{The GFP-based Algorithm for solving~\eqref{P0} }
    \label{alg3}
 \hspace*{0.02in} {\bf Require:} $P_{\max }^{\text{B}}$, $P_{\max }^{\text{R}}$, $\mathbf{G}$, $\mathbf{H}_{\operatorname{B},k},{{\mathbf{H}}_{\text{R},k}}, {{\Gamma }_{k}}, \forall k$, ${{\mathbf{h}}^p_{\operatorname{R},e}}, {{\mathbf{h}}^p_{\operatorname{B},e}}, \forall p$, ${{\Gamma }_{E}}$, ${{J}_{\max }}$, $\zeta >0$. \\
	\hspace*{0.02in} {\bf Ensure:} enhanced solution $\mathbf{x}^{*}$, $\left\{ \mathbf{w}_{p}^{*} \right\}_{p=1}^{P}$,  $\left\{ \mathbf{v}_{k}^{*} \right\}_{k=1}^{K}$ and $\boldsymbol{\vartheta}^{*}$.\\
  \vspace{-0.4cm}
	\begin{algorithmic}[1]		
\State \textbf{Initialize} ${{\mathbf{v}}_k^{\left( 0 \right)}}, \forall k$, ${{\mathbf{x}}^{\left( 0 \right)}}$, ${\boldsymbol{\vartheta }^{(0)}}$ randomly, $j=1$; \\
\textbf{While} $j\le {{J}_{\max }}$ and $\left\| u_p^{(j)}-u_p^{(j-1)} \right\|\ge \zeta $ \textbf{do} \\
\ \ Fixed $\mathbf{x}^{\left( j-1\right)}$, ${{\mathbf{v}}_k^{\left( j-1 \right)}}, \forall k$ and $\boldsymbol{\vartheta}^{\left( j-1\right)}$, update $\mathbf{w}_{p}^{\left( j \right)},\forall p$ by solving~\eqref{wwi}; \\
\ \ Fixed $\mathbf{w}_{p}^{\left( j \right)},\forall p$, $\mathbf{x}^{\left( j-1\right)}$ and $\boldsymbol{\vartheta}^{\left( j-1\right)}$, update ${{\mathbf{v}}_k^{\left( j \right)}}, \forall k$ by solving~\eqref{31}; \\
\ \ Fixed ${{\mathbf{v}}_k^{\left( j \right)}}, \forall k$, $\mathbf{w}_{p}^{\left( j \right)},\forall p$ and $\boldsymbol{\vartheta }^{\left( j-1 \right)}$, update ${{\mathbf{x}}^{(j)}}$ by solving~\eqref{px}; \\
\ \ Fixed ${{\mathbf{x}}^{(j)}}$, ${{\mathbf{v}}_k^{\left( j \right)}}, \forall k$, $\mathbf{w}_{p}^{\left( j \right)},\forall p$, update ${\boldsymbol{\vartheta }^{(j)}}$ using Algorithm~\ref{alg2}; \\
\ \ After obtaining the optimized CRC ${\boldsymbol{\vartheta }^{(j)}}$, the DRC ${\boldsymbol{\hat \vartheta }^{*(j)}}$ is solved through~\eqref{1a},~\eqref{2a} \\
\ \ \textbf{if} $\underset{\theta^p,{\varphi}^p}{\mathop{\min }}\text{SINR}_{p}\left(\boldsymbol{\hat \vartheta}^{*(j)}\right) \!\geq\! \underset{\theta^p,{\varphi}^p}{\mathop{\min }}\text{SINR}_{p}\left(\boldsymbol{\hat \vartheta}^{*(j-1)}\right)$
 \\~~~~ \textbf{then}~ $\boldsymbol{\vartheta}^{(j)}=\boldsymbol{\hat \vartheta}^{*(j)}$\\
~~~~ \textbf{else}~ $\boldsymbol{\vartheta}^{(j)}=\boldsymbol{\hat \vartheta}^{*(j-1)}$\\
~~\textbf{end}\\
\ \ $j=j+1$; \\
\textbf {end while} \\

\textbf{Output} ${{\mathbf{x}}^{*}}={{\mathbf{x}}^{\left( j\right)}}$, $\left\{ \mathbf{w}_{p}^{*} \right\}_{p=1}^{P}=\left\{ \mathbf{w}_{p}^{\left( j \right)} \right\}_{p=1}^{P}$, $\left\{ \mathbf{v}_{k}^{*} \right\}_{k=1}^{K}=\left\{ \mathbf{v}_{k}^{\left( j \right)} \right\}_{k=1}^{K}$ and ${\boldsymbol{\vartheta }}^{*}={\boldsymbol{\vartheta }}^{\left( j \right)}$.
	\end{algorithmic}
\end{algorithm}

Analogous to~\cite{2023Wang}, we employ the classical propagation path loss model as $35.3+37.6{{\log }_{10}}{{l}_{ab}}\left( \text{dB} \right)$ , where ${{l}_{ab}}$ denotes the distance from $a$ to $b$.
The Rician coefficients are all set as $4\,\,\text{dB}$.
We consider that the transmission distances of the BS-Eve, BS-RIS, BS-$\text{SCU}_k$ channels are identical, that is, ${{l}_\text{BE}}={{l}_\text{BR}}={{l}_{\text{B},k}}=30\,\,\text{m}, \forall k$, and the transmission distances of the RIS-Eve and the RIS-$\text{SCU}_k$ channels are identical, that is, ${{l}_\text{RE}}={{l}_{\text{R},k}}=3\,\,\text{m}, \forall k$.
Finally, the stopping conditions of the proposed optimization procedures are assumed as $\zeta=\zeta_1>{{10}^{-5}}$ and ${{\varpi }_{1}}>{{10}^{-4}}$.

To investigate the performance of the devised scheme, we focus on two design approaches, that may be representative of different amount of knowledge available about the TL,
\textbf{Nominal design}:
where no mismatches are considered at the design stage and only the nominal TL is exploited to design the optimized solutions.
The nominal design corresponds to $\mathcal{I}_{TL}=\left[ {{0}^{\circ }},{{0}^{\circ }} \right]$ (${{0}^{\circ }}$ is the nominal TL).
\textbf{Worst-case design}:
where an uncertainty set for the TL is assumed at the design stage.
Thus, a worst-case optimized solution over the mentioned set is provided.
%Three uncertainty sets for the TL are assumed at the optimization phase, i.e., $\left[ -{{2.5}^{\circ }},{{2.5}^{\circ }} \right]$, $\left[ -{{5}^{\circ }},{{5}^{\circ }} \right]$ and $\left[ -{{10}^{\circ }},{{10}^{\circ }} \right]$.
%Therefore, average optimized solutions over the specified uncertainty sets are offered.

\begin{figure}
\captionsetup{font={footnotesize}}
	\begin{center}
	\includegraphics[width=7.5cm]{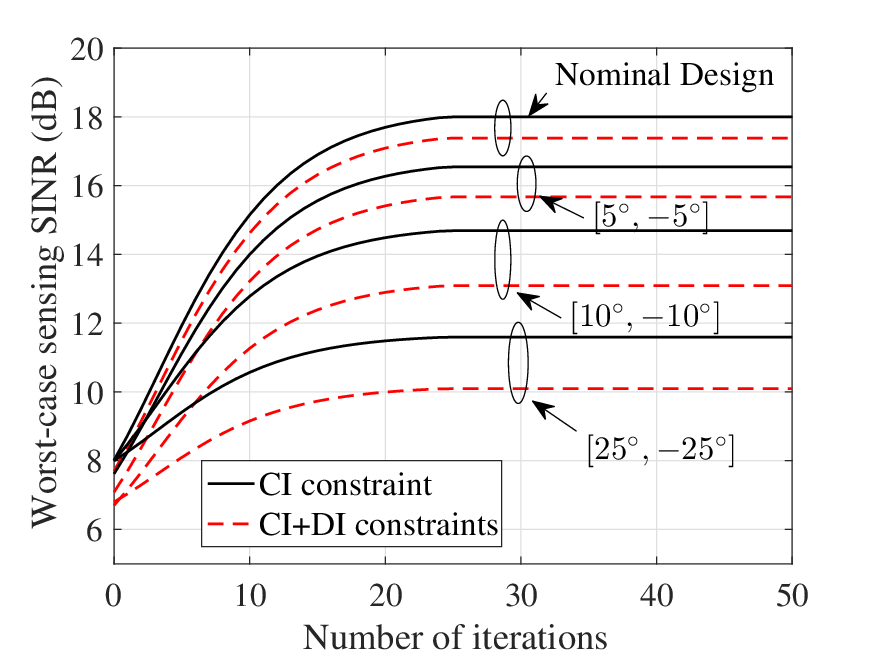}
		\caption{Convergence of devised scheme under the constraints of CI, and CI plus DI over four different uncertainty sets.}
		\label{convergence}
	\end{center}
\end{figure}

The convergence performance of the devised scheme, for several TL uncertainty sets with different sizes, is shown in Fig.~\ref{convergence}, where two cases under the constraints of CI, and CI plus DI are plotted, respectively.
We can observe that the worst-case sensing SINR performance of the devised approach monotonically enhances with the iterations.
The simulation result indicates that the larger the size of the TL uncertainty set, the worse the minimum sensing SINR performance.
%This is a predictable result as a larger set $\mathcal{I}$ suggests a greater uncertainty on the actual TL.
Besides, the proposed optimization algorithm converges fast when the perfect knowledge of TL is available at the BS.
It should be noted that, the worst-case sensing SINR becomes smaller when CI and DI requirements are considered.
{\color{blue}Specifically, when the received symbols at the eavesdropping target are constructed in the destructive area, SCUs decode the received symbols with a lower probability, which indicates the SER performance of the SCUs is deteriorated to some extend when DI-type constraint is taken into account.}

In the following figures, we compare the proposed scheme to the following benchmark techniques:
1) \textbf{No-RIS}:
We assume the hybrid RIS matrix to be a zero matrix, i.e., $\mathbf{\Theta }=\mathbf{0}$, and devise the optimized solutions using the semidefinite relaxation (SDR)-based method in~\cite{2022Su}.
2) \textbf{Optimized passive RIS}:
We assume $A=0$ and $\mathcal{I}_{N_I}^{\mathbb{Q}}=\mathbf{0}$.
The transmit waveform, receive filters and fully-passive RIS coefficients are designed by the technique in~\cite{2022LiuRang}.
3) \textbf{Random passive RIS}:
%The hybrid RIS matrix is considered as $\mathbf{\Theta }=\text{diag}\left\{ {{e}^{j{{\mu }_{1}}}},\cdots ,{{e}^{j{{\mu }_{N_I}}}} \right\}$, where ${\mu }_{n}, n=1,...,N_I$, is randomly generated from $[0,2\pi)$
The discrete reflection coefficients (DRC) are randomly generated according to the uniform distribution~\cite{DRC2024}.
All other variables are designed based on the scheme in Section~\ref{ProbForm}.
4) \textbf{Optimized active RIS}:
We consider $A=N_I$ and $\mathbf{I}_{N_I}-\mathcal{I}_{N_I}^{\mathbb{Q}}=\mathbf{0}$.
The transmit waveform, receive filters and fully-active RIS weights are optimized employing the proposed algorithm.

\subsection{Single Hybrid RIS}\label{single}

\begin{figure}
\captionsetup{font={footnotesize}}
	\begin{center}
	\includegraphics[width=7.5cm]{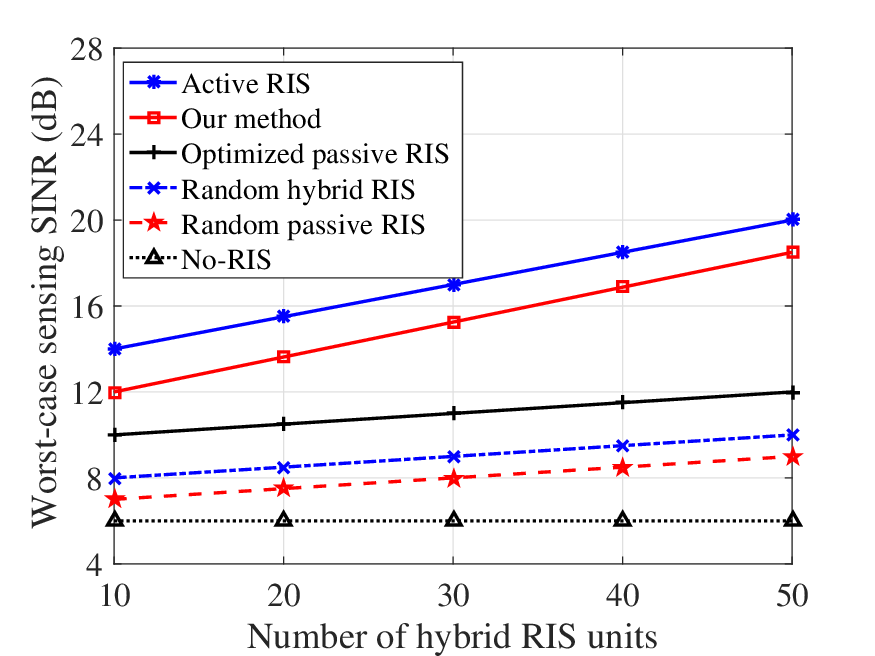}
		\caption{Worst-case sensing SINR versus the number of hybrid RIS units $N_I$, $A=10$.}
		\label{numberRIS}
	\end{center}
\end{figure}

{\color{blue}In this subsection, we show simulation results on the secure ISAC performance versus various parameters in the single hybrid RIS scenario.
The uncertainty set for the TL is assumed as $\left[ -{{10}^{\circ }},{{10}^{\circ }} \right]$.}
Fig.~\ref{numberRIS} illustrates the worst-case sensing SINR versus the number of RIS units $N_I$ of the proposed method and some benchmark approaches, setting $A=10$.
Obviously, our method makes significant performance improvement over the optimized passive RIS and random RIS baselines under the total power budget constraint.
Furthermore, the sensing SINR performance of the devised method is lower than that of the active RIS.
That is because active reflecting units with power amplifiers in hybrid RIS are capable of exploring higher performance enhancement than passive RIS units.
Furthermore, as $N_I$ increases, the worst-case SINR of our method and the optimized passive RIS increases, and the proposed method enhances faster than passive RIS baseline.
In particular, when $N_I=20$, the achieved SINR of Algorithm 3 is about 76.7\% greater than that of optimized passive RIS, and when $ N_I=50$, the SINR improvement is 110.4\% greater.
The phenomenon illustrates that the proposed hybrid RIS strategy strikes a balance between hardware cost and system performance.

\begin{figure}
\captionsetup{font={footnotesize}}
	\begin{center}
	\includegraphics[width=7.5cm]{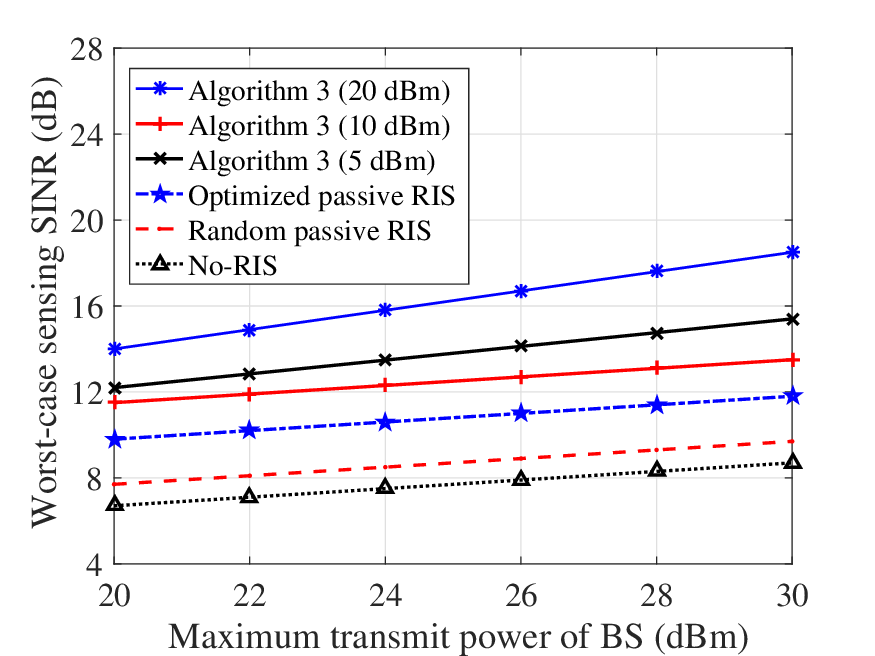}
		\caption{Worst-case sensing SINR versus the maximum power budget of $P_{\max }^{\text{B}}$ for $P_{\max }^{\text{R}}\in \left\{ 5\,\,\text{dBm}, 10\,\,\text{dBm},20\,\,\text{dBm} \right\}$.}
		\label{powerBS}
	\end{center}
\end{figure}

Fig.~\ref{powerBS} shows the worst-case sensing SINR versus the transmit power budget of BS $P_{\max }^{\text{B}}$ for the devised scheme and some benchmark approaches, considering the power consumption of the hybrid RIS $P_{\max }^{\text{R}}\in \left\{5\,\,\text{dBm}, 10\,\,\text{dBm}, 20\,\,\text{dBm}\right\}$.
As expected, all methods are listed in ascending order of the SINR performance: No-RIS, random passive RIS, optimized passive RIS, Algorithm 3 ($P_{\max }^{\text{R}}\in \left\{ 5\, \,\text{dBm},10 \,\,\text{dBm}, 20\,\,\text{dBm} \right\}$).
As $P_{\max }^{\text{B}}$ increases, the sensing SINR performance of all approaches enhances.
Furthermore, the sensing performance of the devised algorithm improves with increasing $P_{\max }^{\text{B}}$.
In particular, the proposed algorithm with $P_{\max}^{\text{R}}=5\,\,\text{dBm}$ and $P_{\max}^{\text{R}}=10\,\,\text{dBm}$ have almost the same sensing performance when $P_{\max }^{\text{B}}$ is low.
The performance gap between them becomes greater with the increasing $P_{\max }^{\text{B}}$.

\begin{figure}
\captionsetup{font={footnotesize}}
	\begin{center}
	\includegraphics[width=7.5cm]{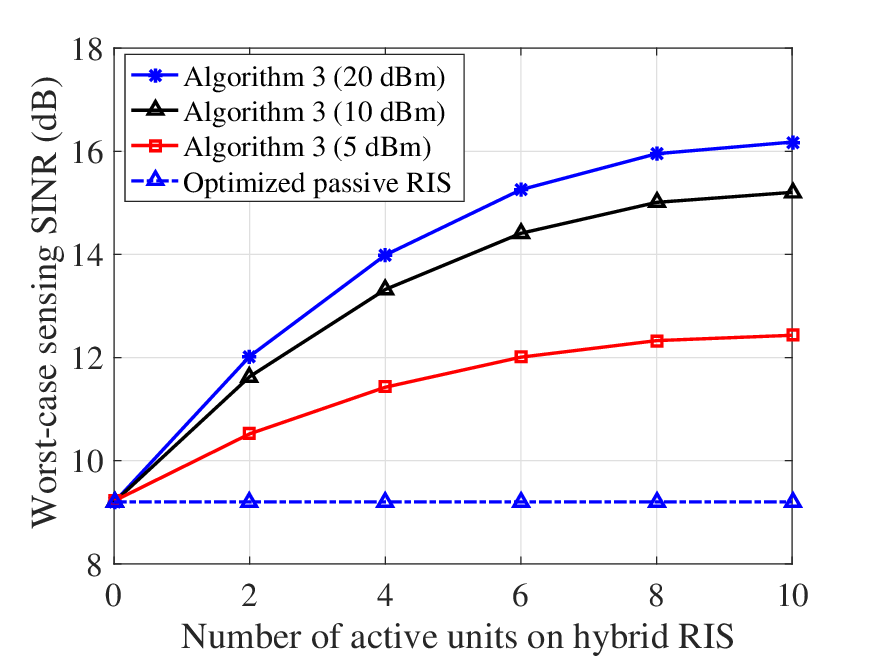}
		\caption{Worst-case sensing SINR versus the number of active units on hybrid RIS for $P_{\max }^{\text{R}}\in \left\{ 5\,\,\text{dBm}, 10\,\,\text{dBm},20\,\,\text{dBm} \right\}$.}
		\label{activeHR}
	\end{center}
\end{figure}

Fig.~\ref{activeHR} illustrates the worst-case sensing SINR versus the number of active units on the hybrid RIS $A$ of the proposed method for $P_{\max }^{\text{R}}\in \left\{ 5\,\,\text{dBm},10\,\,\text{dBm}, 20\,\,\text{dBm} \right\}$, and optimized passive RIS baseline is also considered as a benchmark.
When $A=0$, the proposed approach is reduced to the fully-passive RIS, thus they have the similar sensing performance.
In addition, the sensing SINR performance of our approach improves as $A$ increases.
Specifically, when the number of active units is low, increasing $A$ enhances the sensing performance in a larger proportion.
For Algorithm 3 ($P_{\max }^{\text{R}}=20\,\,\text{dBm}$), the sensing SINR of $A=1$ is 17.5\% greater than that of $A=0$, while the sensing SINR of $A=10$ is only 2.1\% greater than that of $A=9$.
The result indicates that the hybrid RIS strategy dramatically enhances the sensing performance by introducing a small number of active units to passive RIS.
The phenomenon also demonstrates the effectiveness of deploying the optimized hybrid RIS in secure ISAC systems.

\begin{figure}
\captionsetup{font={footnotesize}}
	\begin{center}
		\includegraphics[width=7.5cm]{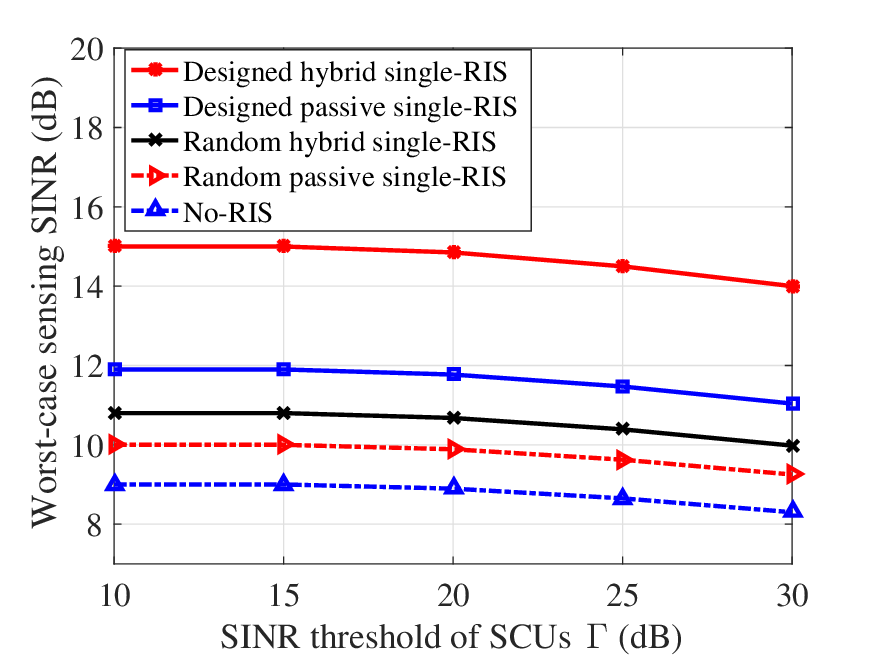}
		\caption{Worst-case sensing SINR versus SCU's SINR with various optimization approaches.}
		\label{comthreshold}
	\end{center}
\end{figure}

\begin{figure}
\captionsetup{font={footnotesize}}
	\begin{center}
		\includegraphics[width=7.5cm]{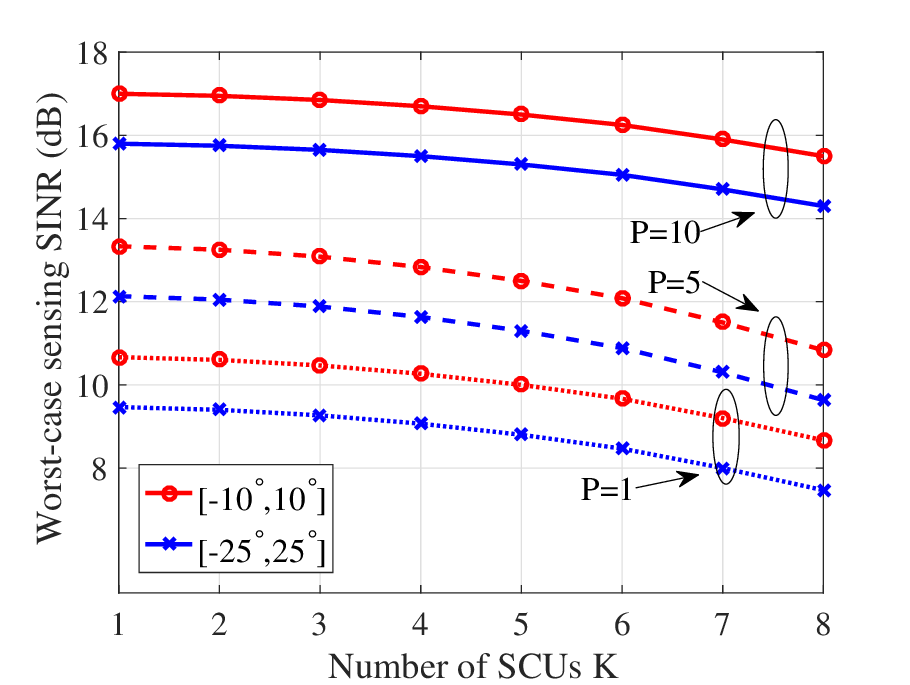}
		\caption{Worst-case sensing SINR versus the number of SCUs with various filter array size.}
		\label{uernumber}
	\end{center}
\end{figure}

Fig.~\ref{comthreshold} depicts {\color{blue}the worst-case sensing SINR} versus the communication QoS constraint ${{\Gamma }_{k}}$.
Clearly, the increase of ${{\Gamma }_{k}}$ has almost no effect on the target sensing capability if ${{\Gamma }_{k}}$ is chosen within a reasonable region, for example, 10\,\,dB-20\,\,dB, for practical communication application scenarios.
This is reasonable behavior because the designed hybrid RIS-aided ISAC system ensures high quality transmission requirement as well as easily satisfies the required QoS constraints, when the objective is to optimize sensing SINR.
The tradeoff between sensing capability and high communication QoS is observed when ${{\Gamma }_{k}}=30\,\,\text{dB}$.
Fig.~\ref{uernumber} displays the worst-case sensing SINR versus the number of SCUs with various number of receive filters $P$, setting two TL uncertainty sets, which exhibits the performance tradeoff between target sensing and multi-user communications.
The result also indicates that worst-case sensing SINR improves as the size of the filter array $P$ increases.
The result verifies the expectation that increasing $P$ (with the TL set fixed) is beneficial for improving the target sensing capability.

\begin{figure}
\captionsetup{font={footnotesize}}
	\begin{center}
\includegraphics[width=7.5cm]{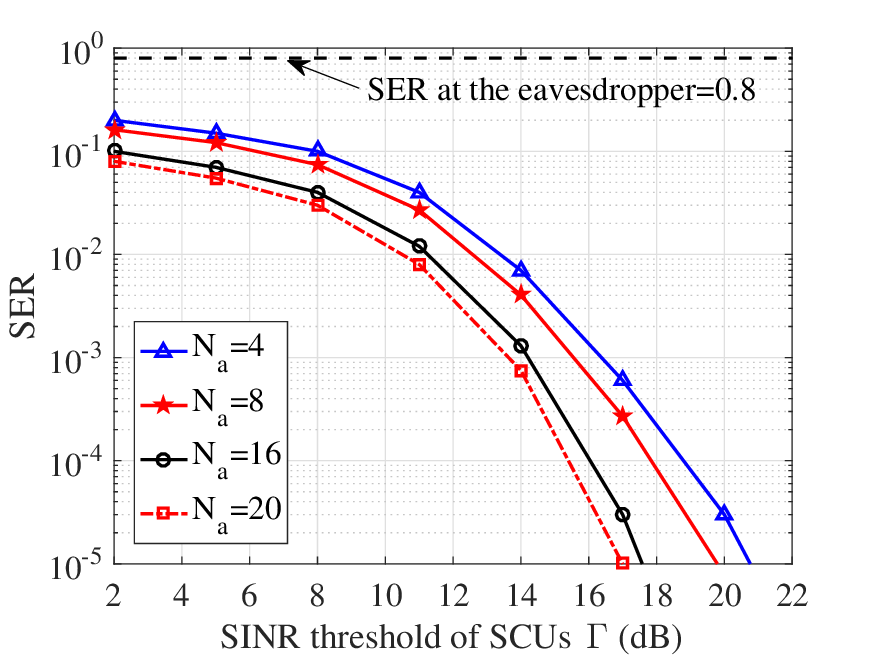}
		\caption{SER of SCU versus the communication QoS constraint ${{\Gamma }_{k}}$ with various number of BS antennas.}
		\label{SER}
	\end{center}
\end{figure}

In Fig.~\ref{SER}, the average SER of all SCUs versus the communication QoS constraint ${{\Gamma }_{k}}$ is shown for different optimization schemes, when the BS is equipped with various number of antennas.
As we can see, more antennas are capable of increasing spatial DoFs to enhance the waveform diversity and to obtain larger energy-focused beamforming gains.
It can be observed that the average SER of all SCUs reduces with increasing ${{\Gamma }_{k}}$.
This is due to the fact that increasing ${{\Gamma }_{k}}$ improves the threshold distance from the received signal to the decision boundary.
Conversely, the SER of the Eve is about 0.8.
That is because the received signals at the Eve will be sent into the destructive area when applying DI-type design.
Therefore, our method is capable of securing the private information while ensuring the quality of reception at each SCU.

{\color{blue}\subsection{Multiple Hybrid RIS}\label{Multiple}

This subsection provides simulation results on the secure ISAC performance versus different parameters in the multiple hybrid RIS scenario.
Fig.~\ref{multiRISpower} shows the minimum sensing SINR versus the maximum power budget of BS.
The single-RIS ISAC with random phase-shifts has a superior performance than the multi-RIS with random phase-shift because, when the phase-shifts are not well designed, multiple RIS employments result in massive interference from clutter scatterers exceeding the useful signal gain from the eavesdropping target.
In addition, the multi-RIS ISAC with the optimized phase-shift undoubtedly offers the greatest sensing performance for the various power budgets of BS owing to the higher indirect channel gain.
Fig.~\ref{multiRISpower1} exhibits the worst-case sensing SINR w.r.t. the power budget of active RIS units.
It is interesting to see that the increase in the transmit power of active RIS units brings the noticeable enhancement of sensing SINR for the multi-RIS ISAC with the optimal phase-shift compared with the non-RIS and single-RIS cases.
For multi-RIS ISAC with random phase-shifts, an increase in the active RIS power may lead to a deterioration of sensing SINR, thus directly underlining the significance of phase-shift optimization design.
Furthermore, it should be noted that in the considered RIS-aided ISAC system, an active unit with a larger amplitude does not always result in a performance enhancement since it also amplifies inter-symbol interference.

\begin{figure}
\captionsetup{font={footnotesize}}
	\begin{center}
\includegraphics[width=7.5cm]{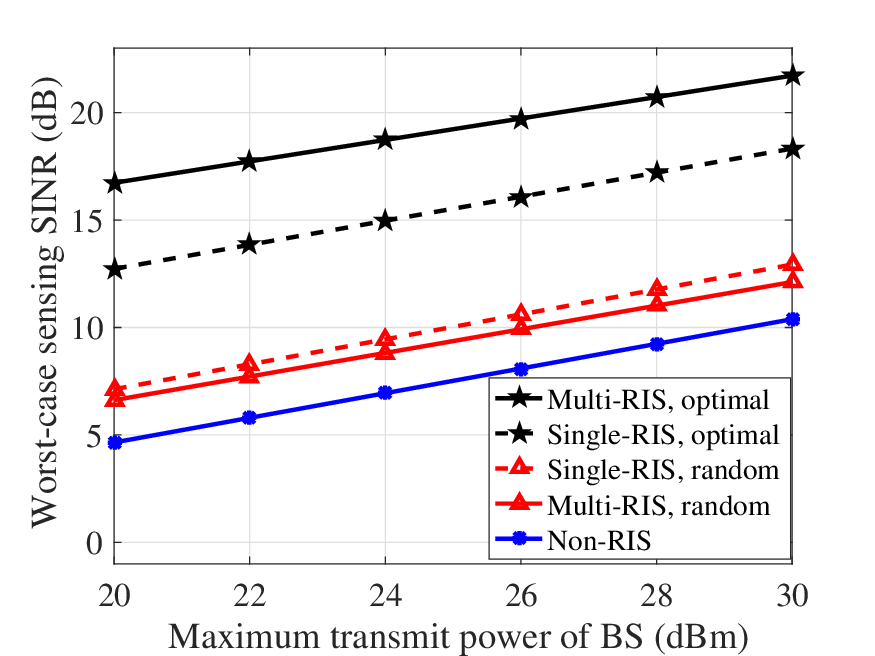}
		\caption{Worst-case sensing SINR versus the maximum power budget of BS for multi-RIS and single-RIS ISAC cases.}
		\label{multiRISpower}
	\end{center}
\end{figure}

\begin{figure}
\captionsetup{font={footnotesize}}
	\begin{center}
\includegraphics[width=7.5cm]{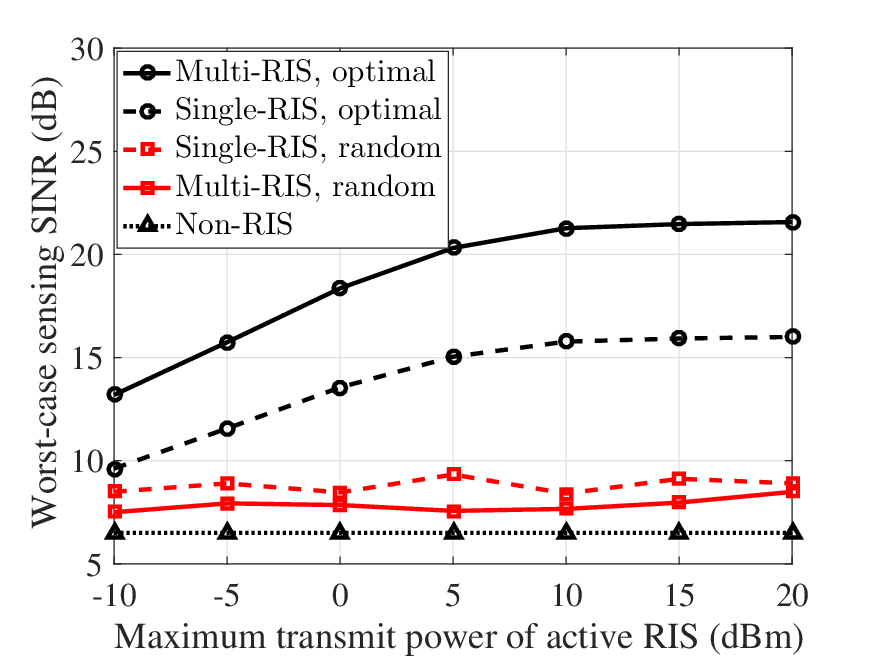}
		\caption{Worst-case sensing SINR versus the maximum power budget of active RIS for multi-RIS and single-RIS ISAC cases.}
		\label{multiRISpower1}
	\end{center}
\end{figure}

In Fig.~\ref{TLuncertainty}, we analyze the influence of correlation between the SCU and Eve LoS path in the sensing performance with different angular interval of TL uncertainty, when the angle difference from the SCU to Eve (i.e., $\Delta\phi$) varies from ${5}^{^\circ }$ to ${25}^{^\circ }$.
It reveals the tradeoff between sensing performance and TL uncertainty.
Furthermore, it is promising to observe that despite the performance loss owing to the CSI errors, the multiple hybrid RIS still offers remarkable performance improvement w.r.t. the multiple passive RIS and non-RIS cases, especially when the angle difference $\Delta\phi$ is low.
Besides, in Fig.~\ref{CIDItradeoff}, one SCU is taken as a reference to assess the SER performance of the Eve versus angle difference between the Eve and SCU.
The uncertainty set for the TL is considered as $\left[ -{{20}^{\circ }},{{20}^{\circ }} \right]$.
It can be observed that Eve decode probability converges to 0.8 with the increasing angular difference when only CI-type requirement is considered.
Furthermore, The SER at the Eve increases noticeably when the DI-type requirement is taken into consideration, which is close to 1 when the angle difference increases.
Therefore, it demonstrates that the employment of DI-type scheme prevents the Eve from wiretapping communication information efficiently.
Moreover, in ISAC system, the multi-RIS-aided system with designed phase-shifts achieves the better secure communication performance compared with the single-RIS-aided system.

\begin{figure}
\captionsetup{font={footnotesize}}
	\begin{center}
\includegraphics[width=7.5cm]{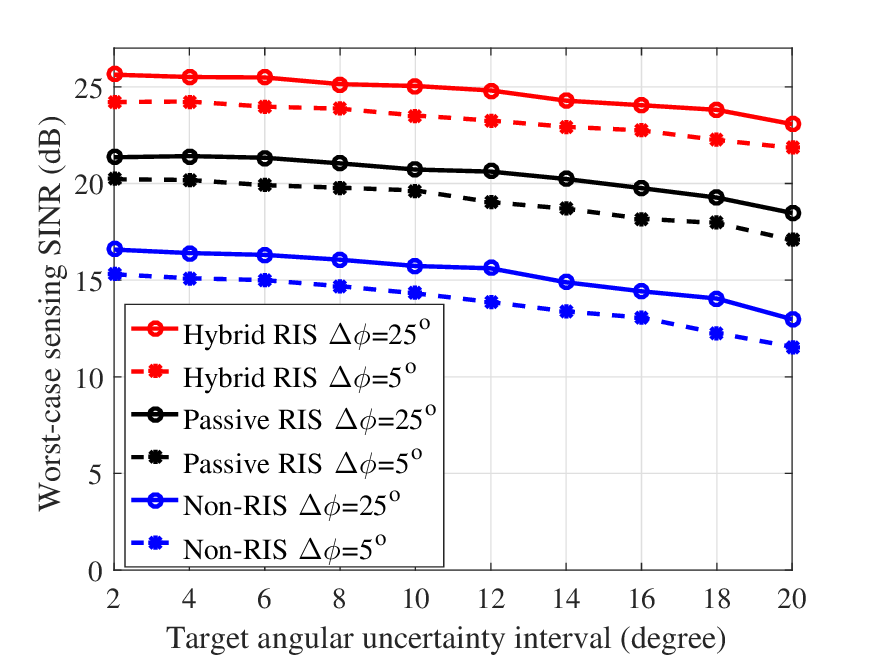}
		\caption{Worst-case sensing SINR versus angular interval of target location uncertainty for hybrid RIS,  passive RIS and non-RIS ISAC cases.}
		\label{TLuncertainty}
	\end{center}
\end{figure}

\begin{figure}
\captionsetup{font={footnotesize}}
	\begin{center}
\includegraphics[width=7.5cm]{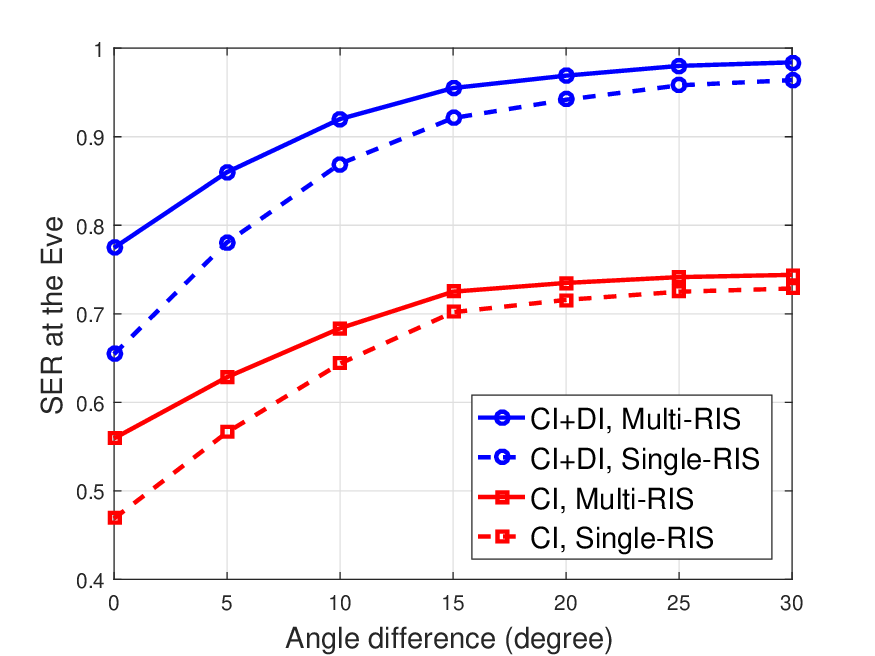}
		\caption{SER at the Eve versus the angle difference between the SCU and Eve with and without DI constraint for multi-RIS and single-RIS ISAC cases.}
		\label{CIDItradeoff}
	\end{center}
\end{figure}

To further verify the performance of the proposed DRC scheme, we simulate the worst-case sensing SINR under the following benchmark approaches.
Specifically, the SDR method along with Gaussian randomization (GR) is adopted in~\cite{DRC2025} to handle the DRC optimization problem.
Fig.~\ref{DRIS} displays the sensing performance versus the resolution of DRC $b$, where the transmit power budgets of BS and active RIS units are $P^\text{B}=20 \,\,\text{dBm}$, and $P^\text{R}=20 \,\,\text{dBm}$ and the number of active reflectors is $A=4$.
Upon improving the resolution of DRC, it can be found that the sensing SINR under the DRC likewise experiences a slow augmentation, eventually converging towards the sensing SINR under the CRC.
Moreover, the sensing SINR under the proposed scheme is higher than that under the Gaussian randomization scheme.
This is because the method in~\cite{DRC2025} exploits RIS to improve the channel gain, whereas here, the proposed scheme is to reduce the communication-sensing mutual interference.
Besides, increasing the resolution of DRC fails to improve system performance under the random DRC (RDRC) benchmark.
Therefore, to manifest the effectiveness of discrete RIS, careful devise of their DRC is indispensable; otherwise, it is possible to unintentionally amplify the mutual interference channel.

\begin{figure}
\captionsetup{font={footnotesize}}
	\begin{center}
\includegraphics[width=7.5cm]{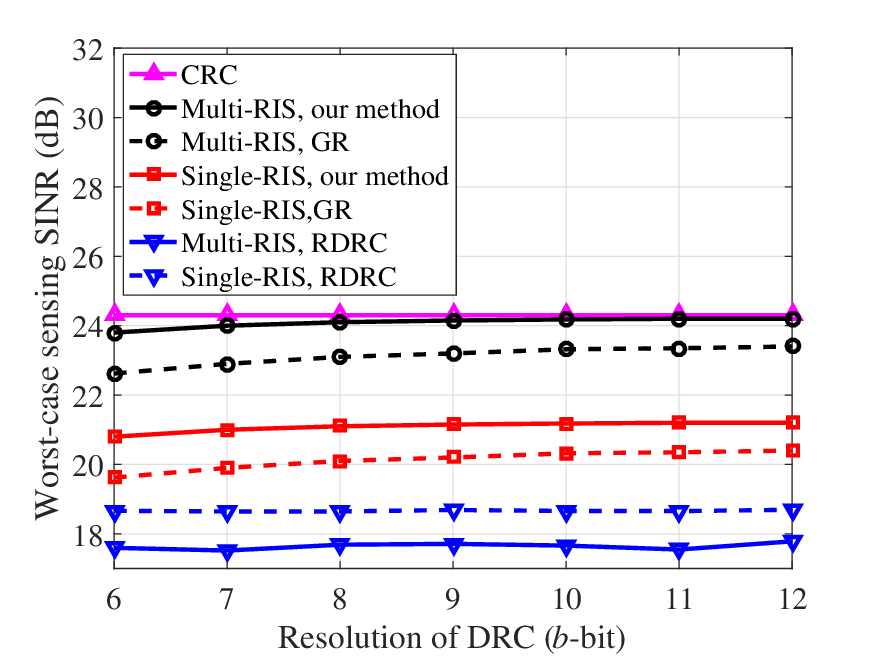}
		\caption{Worst-case sensing SINR versus the resolution of DRC $b$.}
		\label{DRIS}
	\end{center}
\end{figure}

}

\section{Conclusions}\label{Conclusion}

This paper developed an efficient optimization technique for securing multiple hybrid RIS-aided ISAC systems, {\color{blue}which involved the joint design of the transmit signal and receive filters at the BS, the receive beamformers at each SCU, together with the weights at the hybrid RIS.}
The multiple hybrid RIS-aided ISAC design was formulated as a high-dimensional optimization problem to maximize {\color{blue}the worst-case sensing SINR all possible TLs while ensuring the CI-type communication QoS and DI-type security requirements.}
To address the non-convexity of the joint design problem, an alternating optimization procedure {\color{blue}based on GFP, PDD and PCCP techniques} was developed.
The simulation results verified that the devised multiple hybrid RIS scheme can realize more excellent performance than the state-of-the-art benchmark methods.
Moreover, the effectiveness of multiple hybrid RIS for enhancing the ISAC performance was demonstrated and the capability enhancement enlarged with the increasing total power consumption of the system and number of hybrid RIS reflection units.

\begin{appendices}
\section{} \label{3f}

{\color{blue}Using the transformation $\mathbf{a}_{s,e}^{pH}\mathbf{\Theta }_s={\boldsymbol{\vartheta }^{T}_s}\text{diag}\left\{ \mathbf{a}_{s,e}^{pH} \right\}, \forall s$,
the terms $\mathbf{H}^p_{\text{e}}(\boldsymbol{\vartheta },\boldsymbol{\vartheta_1 })$, ${\mathbf{H}}_\text{c}(\boldsymbol{\vartheta },\boldsymbol{\vartheta_1 })$ and ${{\mathbf{\hat{H}}}_{e}^p}(\boldsymbol{\vartheta })$ in the function $g_p$ in~\eqref{aaa1} are redefined as
\begin{equation}
{\mathbf{\bar{H}}}^p_{\text{e}}\mathbf{\bar{T}}\bar{\mathbf{P}}_0\mathbf{x}=\left( \mathbf{I}_{M}\otimes \mathbf{H}_{\mathrm{eg}}^{pH}\boldsymbol{\vartheta}^{*}\boldsymbol{\vartheta}_1^T\mathbf{H}_{\mathrm{eg}}^{p}\right) \mathbf{\bar{T}}{\mathbf{\bar P}}_0\mathbf{x},
\end{equation}
\begin{equation}
{\mathbf{\bar{H}}}_{\text{c}}\mathbf{\bar{T}}\bar{\mathbf{P}}_1\mathbf{x}=\left( \mathbf{I}_{M}\otimes \mathbf{H}_{\mathrm{cg}}^{H}\boldsymbol{\vartheta}^{*}\boldsymbol{\vartheta}_1^T\mathbf{H}_{\mathrm{cg}}\right) \mathbf{\bar{T}}\bar{\mathbf{P}}_1\mathbf{x},
\end{equation}
\begin{equation}
{{\mathbf{\tilde{H}}}_{e}^{pH}}\mathbf{w}_p=\left( \mathbf{I}_{M}\otimes \mathbf{a}_{\text{B},e}^p\boldsymbol{\vartheta}^T{{\mathcal{\hat{I}}}_{{{N}_{I}}}}^{{\mathbb{Q}}}\mathbf{\hat{H}}_{\mathrm{re}}^{p}\right) \mathbf{w}_p,
\end{equation}
where $\mathbf{H}_{\mathrm{eg}}^{p}=\left[\mathbf{a}_{\mathrm{B},e}^{pH},\text{diag}\left\{ \mathbf{a}_{1,e}^{pH} \right\}\mathbf{G}_1,\cdots,\text{diag}\left\{ \mathbf{a}_{S,e}^{pH} \right\}\mathbf{G}_S\right]^T$,
$\mathbf{H}_{\mathrm{cg}}=\left[\mathbf{a}_{\mathrm{B},c}^{H},\text{diag}\left\{ \mathbf{a}_{\text{1},c}^{H} \right\}\mathbf{G}_1,\cdots,\text{diag}\left\{ \mathbf{a}_{S,c}^{H} \right\}\mathbf{G}_S\right]^T$,
$\mathbf{\hat{H}}_{\mathrm{re}}^{p}=\left[\mathbf{0},\text{diag}\left\{ \mathbf{a}_{\text{1},e}^{pH} \right\},\cdots,\text{diag}\left\{ \mathbf{a}_{S,e}^{pH} \right\}\right]^T$ and
$${{\mathcal{\hat{I}}}_{{{N}_{I}}}}^{{\mathbb{Q}}}=\begin{bmatrix}
	\mathbf{0} & \cdots & \mathbf{0} &\mathbf{0}\\
	\mathbf{0} & \mathcal{I}_{N_I}^{\mathbb{Q}} & \cdots & \mathbf{0} \\
	\vdots & \vdots & \ddots & \vdots \\
	\mathbf{0} & \mathbf{0} & \cdots & \mathcal{I}_{N_I}^{\mathbb{Q}}
\end{bmatrix}.$$
Therefore, by utilizing the property of Kronecker product:
\begin{equation}
\mathrm{vec}\{\mathbf{ABC}\}=(\mathbf{C}^T\otimes\mathbf{A})\mathrm{vec}\{\mathbf{B}\},
\end{equation}
we can extract the variable $\boldsymbol{\vartheta}$ from the term ${\mathbf{\bar{H}}}^p_{\text{e}}\mathbf{\bar{T}}\bar{\mathbf{P}}_0\mathbf{x}$,
${\mathbf{\bar{H}}}_{\text{c}}\mathbf{\bar{T}}\bar{\mathbf{P}}_1\mathbf{x}$ and
${{\mathbf{\tilde{H}}}_{e}^{pH}}\mathbf{w}_p$ based on the following transformations
\begin{equation}\label{aa1}
{\mathbf{\bar{H}}}^p_{\text{e}}\mathbf{\bar{T}}\bar{\mathbf{P}}_0\mathbf{x}={{\left( {{\mathbf{H}}_{\text{eg}}^p}{{\mathbf{X}}_{0}} \right)}^{T}}\otimes \mathbf{H}_{\text{eg}}^{pH}\text{vec}\left\{ {\boldsymbol{\vartheta }^{*}}{\boldsymbol{\vartheta }_1^{T}} \right\},
\end{equation}
\begin{equation}\label{aa2}
{\mathbf{\bar{H}}}_{\text{c}}\mathbf{\bar{T}}\bar{\mathbf{P}}_1\mathbf{x}={{\left( {{\mathbf{H}}_{\text{cg}}}{{\mathbf{X}}_{1}} \right)}^{T}}\otimes \mathbf{H}_{\text{cg}}^{H}\text{vec}\left\{ {\boldsymbol{\vartheta }^{*}}{\boldsymbol{\vartheta }_1^{T}} \right\},
\end{equation}
\begin{equation}\label{aa3}
{{\mathbf{\tilde{H}}}_{e}^{pH}}\mathbf{w}_p={{\left( \mathbf{a}_{\text{B},e}^{pH}{{\mathbf{W}_p}} \right)}^{T}}\otimes\left( {{\mathcal{\hat{I}}}_{{{N}_{I}}}}^{{\mathbb{Q}}}\mathbf{\hat{H}}_{\mathrm{re}}^{p}\right) ^H\boldsymbol{\vartheta }^{*},
\end{equation}
where $\mathbf{X}_0\in\mathbb{C}^{N_a\times M}$, $\mathbf{X}_1\in\mathbb{C}^{N_a\times M}$ and $\mathbf{W}_p\in\mathbb{C}^{N_a\times M}$ are the reshaped version of $\mathbf{\bar{T}}\bar{\mathbf{P}}_0\mathbf{x}$, $\mathbf{\bar{T}}\bar{\mathbf{P}}_1\mathbf{x}$ and $\mathbf{w}_p$, respectively.
Plugging the results in~\eqref{aa1},~\eqref{aa2} and~\eqref{aa3} into~\eqref{aaa1}, the function $g_p$ in~\eqref{aaa1} are re-formulated as
\begin{equation}\label{refunction}
\begin{aligned}
&h_p\left( \boldsymbol{\vartheta} ,{\boldsymbol{\vartheta }_{1}} \right)=
2{{u}_{p}}\operatorname{Re} \left( \mathbf{w}_{p}^{H}{{\mathbf{D}}_{\text{e},p}}\text{vec}\left\{ {\boldsymbol{\vartheta }}{\boldsymbol{\vartheta }_{1}^{T}} \right\}  \right)   \\
&~~~~~~-u_{p}^{2}\text{vec}\left\{ {\boldsymbol{\vartheta }^{*}}{\boldsymbol{\vartheta }_{1}^{T}} \right\} ^H{{\mathbf{D}}_{\text{c},p}}\text{vec}\left\{ {\boldsymbol{\vartheta }^{*}}{\boldsymbol{\vartheta }_{1}^{T}} \right\} \\
&~~~~~~-u_{p}^{2}\boldsymbol{\vartheta }^{H}{{\mathbf{F}}_{p}}\boldsymbol{\vartheta }
-u_{p}^{2}\sigma _{R}^{2}\mathbf{w}_p^{H}\mathbf{w}_p,
\end{aligned}
\end{equation}
where for conciseness we define\par
\begin{footnotesize}
\begin{subequations}\label{DDF}
\begin{align}
&{{\mathbf{D}}_{\text{e},p}}={{\left( {{\mathbf{H}}_{\text{eg}}^p}{{\mathbf{X}}_{0}} \right)}^{T}}\otimes \mathbf{H}_{\text{eg}}^{pH},  \\
&{{\mathbf{D}}_{\text{c},p}}=\left( {{\left( {{\mathbf{H}}_{\text{cg}}}{{\mathbf{X}}_{1}} \right)}^{T}}\otimes \mathbf{H}_{\text{cg}}^{H}\right) ^H\mathbf{w}_p\mathbf{w}_p^{H} {{\left( {{\mathbf{H}}_{\text{cg}}}{{\mathbf{X}}_{1}} \right)}^{T}}\otimes \mathbf{H}_{\text{cg}}^{H},   \\
&{{\mathbf{F}}_{p}}={{\left( \mathbf{a}_{\text{B},e}^{pH}{{\mathbf{W}_p}}\! \right)\!}^{*}}\!\otimes\!\left( \! {{\mathcal{\hat{I}}}_{{{N}_{I}}}}^{{\mathbb{Q}}}\mathbf{\hat{H}}_{\mathrm{re}}^{p}\!\right)\!*\!
{{\!\left(\! \mathbf{a}_{\text{B},e}^{pH}{{\mathbf{W}_p}} \!\right)\!}^{T}}\!\otimes\! \left( {{\mathcal{\hat{I}}}_{{{N}_{I}}}}^{{\mathbb{Q}}}\mathbf{\hat{H}}_{\mathrm{re}}^{p}\right) ^H.
\end{align}
\end{subequations}
\end{footnotesize}%
Analogously, by defining\par
\begin{footnotesize}
\begin{subequations}\label{hatHk}
\begin{align}
&{\mathbf{\hat H}}_{k}=
\!\left[\!\mathbf{v}_k^H\mathbf{H}_{\mathrm{B},k},\text{diag}\!\left\{\!\mathbf{v}_k^H\mathbf{H}_{1,k}\! \right\}\!\mathbf{G}_1,\!\cdots\!,\!\text{diag}\!\left\{\! \mathbf{v}_k^H\mathbf{H}_{S,k} \!\right\}\!\mathbf{G}_S\!\right]^T, \\
&{\mathbf{\hat H}}_{e}^{p}=\left[\mathbf{h}_{\mathrm{B},e}^{pH},\text{diag}\left\{ \mathbf{h}_{1,e}^{pH} \right\}\mathbf{G}_1,\!\cdots,\!\text{diag}\left\{ \mathbf{h}_{S,e}^{pH} \right\}\mathbf{G}_S\!\right]^T,
\end{align}
\end{subequations}
\end{footnotesize}%
the CI and DI constraints in an explicit expression w.r.t. $\boldsymbol{\vartheta }$ can be denoted respectively as\par
\begin{footnotesize}
\begin{equation}\label{reCI}
\begin{aligned}
&\left| \operatorname{Im}(\boldsymbol{\vartheta}^T\hat{\mathbf{H}}_{k}\mathbf{x}\left(n \right)s_{k}^{*} \right|  \le  \\
&\!\left(\!\operatorname{Re}(\boldsymbol{\vartheta}^T\hat{\mathbf{H}}_{k}\mathbf{x}\left(n \right)s_{k}^{*})
\!-\!\sqrt{{\Gamma }_{k}\!\left(\!\sigma _\text{RIS}^{2}\boldsymbol{\vartheta }^H\bar{\mathbf{\Lambda}}_k\boldsymbol{\vartheta }+\sigma _{k}^2\right)} \!\right)\!\tan \phi_k, \forall k, n,
\end{aligned}
\end{equation}
\end{footnotesize}%
\begin{footnotesize}
\begin{subequations}\label{reDI}
\begin{align}
&\left( \operatorname{Re}\left(\boldsymbol{\vartheta}^T {\mathbf{\hat H}}_{\mathrm{e}}^{p}\mathbf{x}\left(n\right) s_{k}^{*} \right)
-\sqrt{{\Gamma }_{\text{E}}\left(\sigma _\text{RIS}^{2}\boldsymbol{\vartheta }^H\bar{\mathbf{B}}_e^p \boldsymbol{\vartheta}+\sigma _{e}^2\right)}\right)\tan {{\phi }_{\text{e}}} \nonumber \\
&-\operatorname{Im}\left( \boldsymbol{\vartheta}^T {\mathbf{\hat H}}_{\mathrm{e}}^{p}\mathbf{x}\left(n\right) s_{k}^{*}  \right)-{{\eta }_{n,p}}\Upsilon \le 0,\forall n,p,   \\
&\left( \operatorname{Re}\left(\boldsymbol{\vartheta}^T {\mathbf{\hat H}}_{\mathrm{e}}^{p}\mathbf{x}\left(n\right) s_{k}^{*} \right)
-\sqrt{{\Gamma }_{\text{E}}\left(\sigma _\text{RIS}^{2}\boldsymbol{\vartheta }^H\bar{\mathbf{B}}_e^p \boldsymbol{\vartheta}+\sigma _{e}^2\right)}\right)\tan {{\phi }_{\text{e}}} \nonumber  \\
&+\operatorname{Im}\left(\boldsymbol{\vartheta}^T{\mathbf{ \hat H}}_{\mathrm{e}}^{p}\mathbf{x}\left(n\right) s_{k}^{*}  \right)-(1-{{\eta }_{n,p}})\Upsilon \le 0,\forall n,p,
\end{align}
\end{subequations}
\end{footnotesize}%
where
\begin{equation}\label{Lambdak}
\bar{\mathbf{\Lambda}}_k=\begin{bmatrix}
		0 & \cdots & \mathbf{0}&\mathbf{0}\\
		\mathbf{0} & \bar{\mathbf{B}}_{1,k} & \cdots & \mathbf{0} \\
		\vdots & \vdots & \ddots & \vdots \\
		\mathbf{0} & \mathbf{0} & \cdots & \bar{\mathbf{B}}_{S,k}
	\end{bmatrix},
\end{equation}
and
\begin{equation}\label{Bep}
	\bar{\mathbf{B}}_e^p=\begin{bmatrix}
		0 & \cdots & \mathbf{0} &\mathbf{0}\\
		\mathbf{0} & \bar{\mathbf{B}}_{1,e} & \cdots & \mathbf{0} \\
		\vdots & \vdots & \ddots & \vdots \\
		\mathbf{0} & \mathbf{0} & \cdots & \bar{\mathbf{B}}_{S,e}
	\end{bmatrix},
\end{equation}
and ${\mathbf{\bar B}}_{s,k}=\text{diag}\left\{\mathbf{v}_k^H\mathbf{H}_{s,k}\mathcal{I}_{{{N}_{I}}}^{\mathbb{Q}}\right\} \text{diag}\left\{\mathbf{v}_k^H\mathbf{H}_{s,k}\mathcal{I}_{{{N}_{I}}}^{\mathbb{Q}}\right\}^H$ and
${\mathbf{\bar B}}_{s,e}=\text{diag}\left\{\mathbf{h}_{s,e}^{pH}\mathcal{I}_{{{N}_{I}}}^{\mathbb{Q}}\right\} \text{diag}\left\{\mathbf{h}_{s,e}^{pH}\mathcal{I}_{{{N}_{I}}}^{\mathbb{Q}}\right\}^H$.
In addition, the $s$th RIS power constraint in an explicit expression w.r.t. $\boldsymbol{\vartheta }$ is written as~\eqref{PR1}.
\begin{equation}\label{PRs}
\begin{aligned}
&{{P}_{R,s}}=\sigma _{\text{RIS}}^{2}\boldsymbol{\vartheta}^H_s\mathcal{I}_{{{N}_{I}}}^{\mathbb{Q}}\mathcal{I}_{{{N}_{I}}}^{\mathbb{Q}}\boldsymbol{\vartheta}_s+ \\
&\boldsymbol{\vartheta}^H_s\mathcal{I}_{{{N}_{I}}}^{\mathbb{Q}}\text{diag}\!\left(\! {{\mathbf{x}}^{H}}{{\!\left(\! \mathbf{e}_{n}^{T}\otimes \mathbf{G}_s \!\right)\!}^{H}}\!\right)\! \text{diag}\!\left(\! {{\mathbf{x}}^{H}}{{\!\left(\! \mathbf{e}_{n}^{T}\!\otimes\! \mathbf{G}_s\! \right)\!}^{H}}\!\right)\!\mathcal{I}_{{{N}_{I}}}^{\mathbb{Q}}\boldsymbol{\vartheta}_s.
\end{aligned}
\end{equation}
Combining the re-formulations of the function $h_p\left( \boldsymbol{\vartheta} ,{\boldsymbol{\vartheta }_{1}} \right)$ in~\eqref{refunction}, the CI-type communication QoS constraint~\eqref{reCI}, the DI-type security constraints~\eqref{reDI} and the RIS power constraint~\eqref{PRs},
problem~\eqref{P2} can be equivalently expressed as problem~\eqref{Pvartheta}.}

\end{appendices}

\bibliographystyle{IEEEtran}
\bibliography{IEEEabrv,reference}

\begin{thebibliography}{10}
\providecommand{\url}[1]{#1}
\csname url@samestyle\endcsname
\providecommand{\newblock}{\relax}
\providecommand{\bibinfo}[2]{#2}
\providecommand{\BIBentrySTDinterwordspacing}{\spaceskip=0pt\relax}
\providecommand{\BIBentryALTinterwordstretchfactor}{4}
\providecommand{\BIBentryALTinterwordspacing}{\spaceskip=\fontdimen2\font plus
\BIBentryALTinterwordstretchfactor\fontdimen3\font minus
  \fontdimen4\font\relax}
\providecommand{\BIBforeignlanguage}[2]{{%
\expandafter\ifx\csname l@#1\endcsname\relax
\typeout{** WARNING: IEEEtran.bst: No hyphenation pattern has been}%
\typeout{** loaded for the language `#1'. Using the pattern for}%
\typeout{** the default language instead.}%
\else
\language=\csname l@#1\endcsname
\fi
#2}}
\providecommand{\BIBdecl}{\relax}
\BIBdecl

\bibitem{2020Andrew}
A.~Zhang, M.~L. Rahman, X.~Huang, Y.~J. Guo, and R.~W. Heath, ``Perceptive
  mobile network: Cellular networks with radio vision via joint communication
  and radar sensing,'' \emph{IEEE Veh. Technol. Mag.}, vol.~16, no.~2, pp.
  20--30, 2021.

\bibitem{2021Fan}
F.~Liu, Y.~Cui, C.~Masouros, J.~Xu, T.~X. Han, Y.~C. Eldar, and S.~Buzzi,
  ``Integrated sensing and communications: Towards dual-functional wireless
  networks for {6G} and beyond,'' \emph{IEEE J. Sel. Areas Commun.,}, vol.~40,
  no.~6, pp. 1728--1767, 2022.

\bibitem{Chen3}
S.~Fang, G.~Chen, P.~Xiao, K.-K. Wong, and R.~Tafazolli, ``Intelligent omni
  surface-assisted self-interference cancellation for full-duplex {MISO}
  system,'' \emph{IEEE Trans. Wireless Commun.}, vol.~23, no.~3, pp.
  2268--2281, 2024.

\bibitem{Chen2}
Y.~Wen, G.~Chen, S.~Fang, Z.~Chu, P.~Xiao, and R.~Tafazolli,
  ``{STAR-RIS}-assisted-full-duplex jamming design for secure wireless
  communications system,'' \emph{IEEE Trans. Inf. Forensics Security}, pp.
  1--1, 2024.

\bibitem{2023Jia}
H.~Jia, X.~Li, and L.~Ma, ``Physical layer security optimization with
  {Cramer-Rao} bound metric in {ISAC} systems under sensing-specific imperfect
  {CSI} model,'' \emph{IEEE Trans. Veh. Technol.}, pp. 1--13, 2023.

\bibitem{2023Ren}
Z.~Ren, L.~Qiu, J.~Xu, and D.~W.~K. Ng, ``Robust transmit beamforming for
  secure integrated sensing and communication,'' \emph{IEEE Trans. Commun.},
  vol.~71, no.~9, pp. 5549--5564, 2023.

\bibitem{2018Jonathon}
A.~Deligiannis, A.~Daniyan, S.~Lambotharan, and J.~A. Chambers, ``Secrecy rate
  optimizations for {MIMO} communication radar,'' \emph{IEEE Trans. Aerosp.
  Electron. Syst.}, vol.~54, no.~5, pp. 2481--2492, 2018.

\bibitem{2022Su}
N.~Su, F.~Liu, Z.~Wei, Y.~F. Liu, and C.~Masouros, ``Secure dual-functional
  radar-communication transmission: Exploiting interference for resilience
  against target eavesdropping,'' \emph{IEEE Trans. Wireless Commun.}, vol.~18,
  no.~6, pp. 1728--1767, 2023.

\bibitem{2022Xu}
H.~Xu, T.~Yang, K.-K. Wong, and G.~Caire, ``Achievable regions and precoder
  designs for the multiple access wiretap channels with confidential and open
  messages,'' \emph{IEEE J. Sel. Areas Commun.}, vol.~40, no.~5, pp.
  1407--1427, 2022.

\bibitem{2017Batu}
B.~K. Chalise, M.~G. Amin, and B.~Himed, ``Performance tradeoff in a unified
  passive radar and communications system,'' \emph{IEEE Signal Process. Lett.},
  vol.~24, no.~9, pp. 1275--1279, 2017.

\bibitem{2018LiuTSP}
F.~Liu, C.~Masouros, A.~Li, T.~Ratnarajah, and J.~Zhou, ``{MIMO} radar and
  cellular coexistence: A power-efficient approach enabled by interference
  exploitation,'' \emph{IEEE Trans. Signal Process.}, vol.~66, no.~14, pp.
  3681--3695, 2018.

\bibitem{2019Khandaker}
M.~R.~A. Khandaker, C.~Masouros, K.-K. Wong, and S.~Timotheou, ``Secure {SWIPT}
  by exploiting constructive interference and artificial noise,'' \emph{IEEE
  Trans. Wireless Commun.}, vol.~67, no.~2, pp. 1326--1340, 2019.

\bibitem{2021Wu}
Q.~Wu, S.~Zhang, B.~Zheng, C.~You, and R.~Zhang, ``Intelligent reflecting
  surface-aided wireless communications: A tutorial,'' \emph{IEEE Trans.
  Wireless Commun.}, vol.~69, no.~5, pp. 3313--3351, 2021.

\bibitem{2021Wang}
C.~Wang, Z.~Li, T.-X. Zheng, D.~W.~K. Ng, and N.~Al-Dhahir, ``Intelligent
  reflecting surface-aided secure broadcasting in millimeter wave symbiotic
  radio networks,'' \emph{IEEE Trans. Veh. Technol.}, vol.~70, no.~10, pp.
  11\,050--11\,055, 2021.

\bibitem{2023Hua}
M.~Hua, Q.~Wu, W.~Chen, O.~A. Dobre, and A.~Lee~Swindlehurst, ``Secure
  intelligent reflecting surface aided integrated sensing and communication,''
  \emph{IEEE Trans. Wireless Commun.}, pp. 1--1, 2023.

\bibitem{2022LiuRang}
R.~Liu, M.~Li, Y.~Liu, Q.~Wu, and Q.~Liu, ``Joint transmit waveform and passive
  beamforming design for {RIS}-aided {DFRC} systems,'' \emph{IEEE J. Sel.
  Topics Signal Process.}, vol.~16, no.~5, pp. 995--1010, 2022.

\bibitem{2022Wang}
F.~Wang, H.~Li, and J.~Fang, ``Joint active and passive beamforming for
  {IRS}-assisted radar,'' \emph{IEEE Signal Process. Lett.}, vol.~29, pp.
  349--353, 2022.

\bibitem{2022Huang}
N.~Huang, T.~Wang, Y.~Wu, Q.~Wu, and T.~Q.~S. Quek, ``Integrated sensing and
  communication assisted mobile edge computing: An energy-efficient design via
  intelligent reflecting surface,'' \emph{IEEE Signal Process. Lett.}, vol.~11,
  no.~10, pp. 2085--2089, 2022.

\bibitem{Jiangzhou2024}
Z.~Yu, H.~Ren, C.~Pan, G.~Zhou, B.~Wang, M.~Dong, and J.~Wang, ``Active
  {RIS}-aided {ISAC} systems: Beamforming design and performance analysis,''
  \emph{IEEE Trans. Commun.}, vol.~72, no.~3, pp. 1578--1595, 2024.

\bibitem{HybridAP1}
C.~Liao, F.~Wang, G.~Han, Y.~Huang, and V.~K.~N. Lau, ``Beamforming design for
  hybrid active-passive {RIS} assisted integrated sensing and communications,''
  \emph{IEEE Commun. Lett.}, vol.~27, no.~11, pp. 2938--2942, 2023.

\bibitem{HybridAP2}
D.~V.~Q. Rodrigues and T.~Singh, ``Efficient power allocation strategies in
  hybrid active-passive reconfigurable intelligent surfaces,'' \emph{IEEE
  Commun. Lett.}, vol.~28, no.~1, pp. 113--117, 2024.

\bibitem{MEC2024}
H.~Xie, D.~Li, and B.~Gu, ``Exploring hybrid active-passive {RIS}-aided {MEC}
  systems: From the mode-switching perspective,'' \emph{IEEE Trans. Wireless
  Commun.}, pp. 1--1, 2024.

\bibitem{MultiRISImperfect023}
Z.~Chen, J.~Tang, X.~Y. Zhang, Q.~Wu, G.~Chen, and K.-K. Wong, ``Robust hybrid
  beamforming design for {Multi-RIS} assisted {MIMO} system with imperfect
  {CSI},'' \emph{IEEE Trans. Wireless Commun.}, vol.~22, no.~6, pp. 3913--3926,
  2023.

\bibitem{MultiRIS2}
P.~Wang, J.~Fang, X.~Yuan, Z.~Chen, and H.~Li, ``Intelligent reflecting
  surface-assisted millimeter wave communications: Joint active and passive
  precoding design,'' \emph{IEEE Trans. Veh. Technol.}, vol.~69, no.~12, pp.
  14\,960--14\,973, 2020.

\bibitem{2022Shu}
F.~Shu, L.~Yang, X.~Jiang, W.~Cai, W.~Shi, M.~Huang, J.~Wang, and X.~You,
  ``Beamforming and transmit power design for intelligent reconfigurable
  surface-aided secure spatial modulation,'' \emph{IEEE J. Sel. Topics Signal
  Process.}, vol.~16, no.~5, pp. 933--949, 2022.

\bibitem{2021Shu}
F.~Shu, Y.~Teng, J.~Li, M.~Huang, W.~Shi, J.~Li, Y.~Wu, and J.~Wang, ``Enhanced
  secrecy rate maximization for directional modulation networks via {IRS},''
  \emph{IEEE Trans. Commun.}, vol.~69, no.~12, pp. 8388--8401, 2021.

\bibitem{2023Wang}
C.~Wang, C.-C. Wang, Z.~Li, D.~W.~K. Ng, K.-K. Wong, N.~Al-Dhahir, and
  D.~Niyato, ``{STAR-RIS}-enabled secure dual-functional radar-communications:
  Joint waveform and reflective beamforming optimization,'' \emph{IEEE Trans.
  Inf. Forensics Security}, vol.~18, pp. 4577--4592, 2023.

\bibitem{ExtendedTargets2024}
Y.~Wang, M.~Tao, and S.~Sun, ``{Cramer-Rao} bound analysis and beamforming
  design for integrated sensing and communication with extended targets,''
  \emph{IEEE Trans. Wireless Commun.}, vol.~23, no.~11, pp. 15\,987--16\,000,
  2024.

\bibitem{ISACCSI1}
Y.~Zhang, W.~Ni, J.~Wang, W.~Tang, M.~Jia, Y.~C. Eldar, and D.~Niyato, ``Robust
  transceiver design for covert integrated sensing and communications with
  imperfect {CSI},'' \emph{IEEE Trans. Commun.}, pp. 1--1, 2024.

\bibitem{2022Chen}
L.~Chen, Z.~Wang, Y.~Du, Y.~Chen, and F.~R. Yu, ``Generalized transceiver
  beamforming for {DFRC} with {MIMO} radar and {MU-MIMO} communication,''
  \emph{IEEE J. Sel. Areas Commun.}, vol.~40, no.~6, pp. 1795--1808, 2022.

\bibitem{2015Masouros}
C.~Masouros and G.~Zheng, ``Exploiting known interference as green signal power
  for downlink beamforming optimization,'' \emph{IEEE Trans. Signal Process.},
  vol.~63, no.~14, pp. 3628--3640, 2015.

\bibitem{2011Shi}
L.~Shi, P.~Wang, H.~Liu, L.~Xu, and Z.~Bao, ``Radar {HRRP} statistical
  recognition with local factor analysis by automatic {Bayesian} ying-yang
  harmony learning,'' \emph{IEEE Trans. Signal Process.}, vol.~59, no.~2, pp.
  610--617, 2011.

\bibitem{2023Nguyen}
N.~T. Nguyen, V.-D. Nguyen, H.~V. Nguyen, H.~Q. Ngo, S.~Chatzinotas, and
  M.~Juntti, ``Spectral efficiency analysis of hybrid relay-reflecting
  intelligent surface-assisted cell-free massive {MIMO} systems,'' \emph{IEEE
  Trans. Wireless Commun.}, vol.~22, no.~5, pp. 3397--3416, 2023.

\bibitem{2015MaioDe}
D.~Maio, Antonio, Karbasi, S.~Mohammad, Aubry, Augusto, Bastani, and M.~Hasan,
  ``Robust transmit code and receive filter design for extended targets in
  clutter,'' \emph{IEEE Trans. Signal Process.}, vol.~63, no.~8, pp.
  1965--1976, 2015.

\bibitem{2013Cheng}
Y.~Cheng, M.~Pesavento, and A.~Philipp, ``Joint network optimization and
  downlink beamforming for {CoMP} transmissions using mixed integer conic
  programming,'' \emph{IEEE Trans. Signal Process.}, vol.~61, no.~16, pp.
  3972--3987, 2013.

\bibitem{Shi2020}
Q.~Shi and M.~Hong, ``Penalty dual decomposition method for nonsmooth nonconvex
  optimization-part {I}: Algorithms and convergence analysis,'' \emph{IEEE
  Trans. Signal Process.}, vol.~68, pp. 4108--4122, 2020.

\bibitem{2016thomas}
T.~Lipp and S.~Boyd, ``Variations and extension of the convex-concave
  procedure,'' \emph{Optim. Eng.}, vol.~17, no.~2, pp. 263--287, 2016.

\bibitem{DRC2024}
L.~Du, Q.~Liang, Z.~Ma, and P.~Fan, ``Over-the-air interference suppression
  using reconfigurable intelligent surface with discrete reflection
  coefficients {(DRC)} design,'' \emph{IEEE Trans. Veh. Technol.}, vol.~73,
  no.~5, pp. 7310--7315, 2024.

\bibitem{DRC2025}
L.~Du, P.~Fan, Z.~Ma, and Q.~Liang, ``{RIS} assisted radar-communication
  coexistence system with discrete reflection coefficients,'' \emph{IEEE Trans.
  Wireless Commun.}, vol.~24, no.~4, pp. 3014--3028, 2025.

\end{thebibliography}

\end{document}